\documentclass[useAMS,usenatbib,referee]{mn2e}

\usepackage{graphics}
\usepackage{epsfig}
\usepackage{latexsym}
\newcommand{\be}{\begin{equation}}
\newcommand{\ee}{\end{equation}}
\newcommand{\ba}{\begin{eqnarray}}
\newcommand{\ea}{\end{eqnarray}}
\newcommand{\bc}{\begin{center}}
\newcommand{\ec}{\end{center}}

\newcommand{\mnras}{\,{\rm MNRAS}}
\newcommand{\apj}{\,{\rm ApJ}}

\newcommand{\aap}{\,{\rm A\&A}}

\usepackage{graphicx}

\newcommand{\fermi}{{\em Fermi}}
\newcommand{\dg}{$^{\circ}$}

\def\ergscm2 {erg\,s$^{-1}$cm$^{-2}$}

   \newcommand{\ti}{t_{\rm i}}
\newcommand{\Rc}{R_{\rm c}}
  \newcommand{\Rs}{R_{\rm SNR}}
    
\newcommand{\ESNR}{E_{\ssst\rm SNR}}

\newcommand{\s}{\,{\rm s}}      
    
\newcommand{\cm}{\,{\rm cm}}

\newcommand{\ssst}{\scriptscriptstyle}

\newcommand{\etal}{et al.}

\title{Cosmic rays in the surroundings of SNR G35.6-0.4 }

\author[Torres et al.]
{Diego F. Torres$^{1,2}$\thanks{dtorres@ieec.uab.es}, Hui Li$^{3}$, Yang Chen$^{3,4}$, Anal\'ia Cillis$^5$,  \newauthor  Andrea G. Caliandro$^1$,  
Ana Y. Rodr\'iguez-Marrero$^6$ \\
$^1$Instituci\'o Catalana de Recerca i Estudis Avan\c{c}ats (ICREA) Barcelona,Spain \\ $^2$Institut de Ci\`encies de l'Espai (IEEC-CSIC), Campus UAB,  Torre C5, 2a planta, 08193 Barcelona, Spain \\
$^3$Department of Astronomy, Nanjing University, Nanjing 210093, China\\
$^4$Key Laboratory of Modern Astronomy and Astrophysics, Nanjing University, Ministry of Education, China\\
$^5$Instituto de Astronom\'ia y F\'isica del Espacio (CONICET-UBA), CC67, Suc. 28, 1428 Buenos Aires, Argentina \\
$^6$Instituto de F\'isica de Cantabria, Edificio Juan Jord\'a, Avenida de los Castros s/n, 39005 Santander, Cantabria, Spain\\
}

\begin{document}

\date{}

\pagerange{\pageref{firstpage}--\pageref{lastpage}} \pubyear{2008}

\maketitle

\label{firstpage}

\begin{abstract}
HESS J1858+020 is a TeV gamma-ray source that
was reported not to have any clear cataloged counterpart at any wavelength. However, it has been recently proposed that 
this source is indirectly associated with 
the radio source,  re-identified
as a supernova remnant (SNR), G35.6-0.4. The latter is found to be middle-aged ($\sim 30$ kyr) and to have
nearby molecular clouds (MCs). HESS J1858+020 was proposed to be the result of the interaction of protons accelerated 
in the SNR
shell with target ions residing in the clouds. 
The \fermi\ Large Area Telescope (LAT) First Source Catalog does not list any source coincident with the position of HESS J1858+020, but some lie close. Here,
we analyse more than 2 years of data obtained with the \fermi-LAT for the region of interest, and consider whether it is indeed possible that 
the closest LAT source, 1FGL J1857.1+0212c, is related to HESS J1858+020. We conclude it is not, and
we impose upper limits on the GeV emission originating from HESS J1858+020.
Using a simplified 3D model for the cosmic-ray propagation out from the shell of the SNR, we consider whether the interaction between SNR G35.6-0.4 and the MCs nearby could give rise to the TeV emission of HESS J1858+020 without producing a GeV counterpart. If so, the pair of SNR/TeV source with no GeV detection would be reminiscent of other similarly-aged SNRs, such as some of the TeV hotspots near W28, for which cosmic-ray diffusion may be used to explain their multi-frequency phenomenology. However, for HESS J1858+020, we found that although the phase space in principle allows for such GeV--TeV non-correlation to appear, usual and/or observationally constrained values of the parameters (e.g., diffusion coefficients and cloud-SNR likely distances) would disfavor it. 
\end{abstract}


\section{Introduction}

HESS J1858+020 is a weak gamma-ray source { (1.6\% Crab Flux)} that
was reported not to have any clear cataloged counterpart at any wavelength (Aharonian et al. 2008).
{ HESS J1858+020 is 
a nearly point-like source, with a slight extension of $\sim 5$ arcmin along its major axis. Its differential spectral index is 2.2 $\pm$ 0.1.}
The nearby radio source G35.6-0.4 was recently re-identified as a SNR (Green 2009).
HESS J1858+020 lies towards the southern border of this
remnant. Paron \& Giacani (2010) have found, using the $^{13}$CO { (J=1-0)}
line from the Galactic Ring Survey and mid-IR data from the Galactic Legacy Infrared Mid-Plane Survey Extraordinaire (GLIMPSE), that
there is one or several MCs towards the southern border of SNR
G35.6-0.4, likely at the same distance of the remnant (10.5 kpc).
Paron \& Giacani (2010) also provide estimates of the cloud's total molecular mass and density. { They report that the
cloud is composed of two molecular clumps, one over the SNR shell and the other located at the center of HESS J1858+020 with  
a molecular mass and a density of $\sim 5 \times 10^3$ M$_\odot$ and $\sim 500$ cm$^{-3}$, respectively.} 
They proposed, using a simplified analytical approach described in Torres et al. (2003), 
that hadronic gamma-ray emission within the clouds, produced by protons diffusing away from the SNR
G35.6-0.4, is a possible origin of HESS J1858+020. 
In a more recent paper, Paron et al. (2011) give more details about the molecular material, obtained via observations with the
Atacama Submillimeter Telescope Experiment. They 
discovered a young stellar object (YSO), probably a high mass protostar, embedded in the molecular clump, but no
evidence of any molecular outflows which might in principle reveal the presence of a thermal jet capable of generating the observed gamma-rays. Paron et al. (2011)  concluded again that 
the most probable origin for the TeV gamma-ray emission are hadronic interactions between the molecular gas and the cosmic rays accelerated by the shock front of SNR G35.6-0.4.
Here, we focus on a more in-depth analysis of this possibility.

An important ingredient to assess the latter proposal is to have an understanding of the GeV emission (if any), of the region. 
However, the GeV emission at the sky position of HESS J1858+020 has not been studied up to now.
A look at
the \fermi-LAT  First Source Catalog (Abdo et al. 2010a) shows that there is no source at the position of the SNR or the H.E.S.S. source. The \fermi\ source
1FGL J1857.1+0212c is the closest one: it is only at 0.3$^{\rm o}$ from
HESS J1858+020, it is unidentified, 
has { an} error radius of 0.08$^o$, and
its spectral index is 2.31$ \pm$     0.04. 
%with pivot energy at    1246.10 MeV.
In the 1FGL analysis there is no indication that 1FGL J1857.1+0212c is confused.
{ The pulsar PSR B1855+02 lies close (see the map below), 
but no gamma-ray emission has been detected originating from it in the \fermi-LAT data (Abdo et al. 2010b).}
PSR B1855+02 was suggested
as being possibly related with the SNR G35.6-0.4 but the association is
unclear (see Green 2009 for details). The { proximity} of HESS J1858+020 and 1FGL J1857.1+0212c, particularly 
since the point spread function (PSF) of the \fermi-LAT instrument is larger than 
the separation of the two sky coordinates of interest, 
requires a careful analysis in order to evaluate the possible relations among these sources.

The case of middle-aged SNRs interacting with MCs,
with the subsequent production of  gamma-ray sources at GeV and/or TeV energies
has also been analysed, for instance, for W51 (Abdo et al.\ 2009),
W28 (Abdo et al.\ 2010e, Li \& Chen 2010, Ohira et al.\ 2011),
and IC 443 (e.g., Abdo et al. 2010f, Torres et al. 2010, see also Rodriguez Marrero et al. 2008).
However, in these cases there was a GeV detection, except for
two of the four TeV hotspots of the SNR W28, namely HESS J1800-230A and
HESS J1800-230C for which only upper limits were imposed.
These instances of GeV -- TeV non-simultaneous detections around SNRs
are probably the most interesting ones from a cosmic-ray viewpoint (see the discussion in Funk et al. 2008). 
They allow us to explore { an} hadronic origin of the highest-energy gamma-ray emission, through cosmic rays
accelerated by the SNR shock front, which, diffusing away from it, interact with MCs at a certain distance. Given that only the most energetic protons reach the separate clouds at a given time, 
these cases allow us to investigate  the 
diffusion environment in which the cosmic-ray propagation proceeds. 

Section 2 is the observational core of the paper and we present there the analysis of more than 2 years of 
\fermi-LAT data of this region. We put special care to analyse whether a mis-localization of 1FGL J1857.1+0212c could render it coincident in projection with HESS J1828+020 and prove that this is likely not the case. We then impose upper limits to the GeV radiation at
the position of HESS J1858+020. These limits are used in Section 3 to
establish possible scenarios where the constraints (detection at TeV,
non-detection at GeV) could be satisfied. A discussion of these results
and a comparison with the similar case of SNR W28 with nearby MCs
(MCs) is given at the end.

\section{GeV analysis}

We searched for gamma rays emitted by the HESS 
J1858+020 analyzing the publicly available
$\sim28$ months of \fermi-LAT data in the energy range 100 MeV - 100 GeV, from 4 Aug 2008 to 19 November 2010.
The search is performed by means of the binned likelihood spectral analysis,
using the official tool ({\tt gtlike})
released by the \fermi-LAT collaboration (\fermi\ Science Tools {\tt v9r17p0}).
All the data, software, and diffuse models used for this analysis are available from the \fermi\ Science Support Center.\footnote{http://fermi.gsfc.nasa.gov/ssc/}
Events from the ``Pass 6 Diffuse'' class were
selected, i.e. the event class with the greatest purity of gamma rays, having the most
stringent background rejection (see details in Atwood et al. 2009).
The ``Pass 6 v3 Diffuse" instrument response functions (IRFs) were applied in the analysis (Rando 2009).
We selected events with energy $E>$100\,MeV in a square aligned with celestial coordinates,
inscribed inside a circular region of interest (ROI) of 15$^{\circ}$ radius,
centered on the H.E.S.S. source.
The good time intervals are defined such that the ROI
does not go below the gamma-ray-bright Earth limb (defined at 105$^{\circ}$ from the
Zenith angle), and that the source is always inside the LAT field of view, namely in a cone angle of
66$^{\circ}$. Source detection significance is determined using the Test-Statistic (TS) value, $TS = -2(L_0 -
L_1)$, which compares the likelihood ratio of models including an additional source, $L_1$, with the
null-hypothesis of background only, $L_0$ (Mattox et al. 1996).

To apply the likelihood analysis, a spectral-spatial model containing diffuse and
point-like sources was created. Using the 1FGL Catalog we have 93 sources closer than 20$^o$ and 5 closer than 3$^o$ from the HESS J1858+020 position.
For the Galactic diffuse emission we used the spectral-spatial model
``gll\_iem\_v02.fit", which is the one used by the \fermi-LAT collaboration in order to build the
 \fermi-LAT First Source Catalog (Abdo et al. 2010a; 1FGL as referred to hereafter). The
isotropic diffuse emission was modelled by the spectrum
described in the ``isotropic\_iem\_v02.txt" file.
The normalization factors of these two
models were left free in the fitting procedures.
The spectral-spatial model also included all the point-like sources from the 1FGL list
closer than 20$^{\circ}$ to the H.E.S.S. source. Each of those point sources was  modeled with a simple
power-law, with the exceptions of those sources that are known to be pulsars, for which a power-law with an exponential cut-off
was used. The spectral parameters of those sources were set at
the 1FGL values or those from the \fermi-LAT First Pulsar Catalog (Abdo et al. 2010b).
The flux parameter of all the point-like sources closer than 3$^\circ$ to HESS 1858+020 were left free in the likelihood fit, except for the closest one, to which we pay special attention below. 

\begin{figure}
\centering
 \includegraphics[width=0.6\textwidth, angle=0]{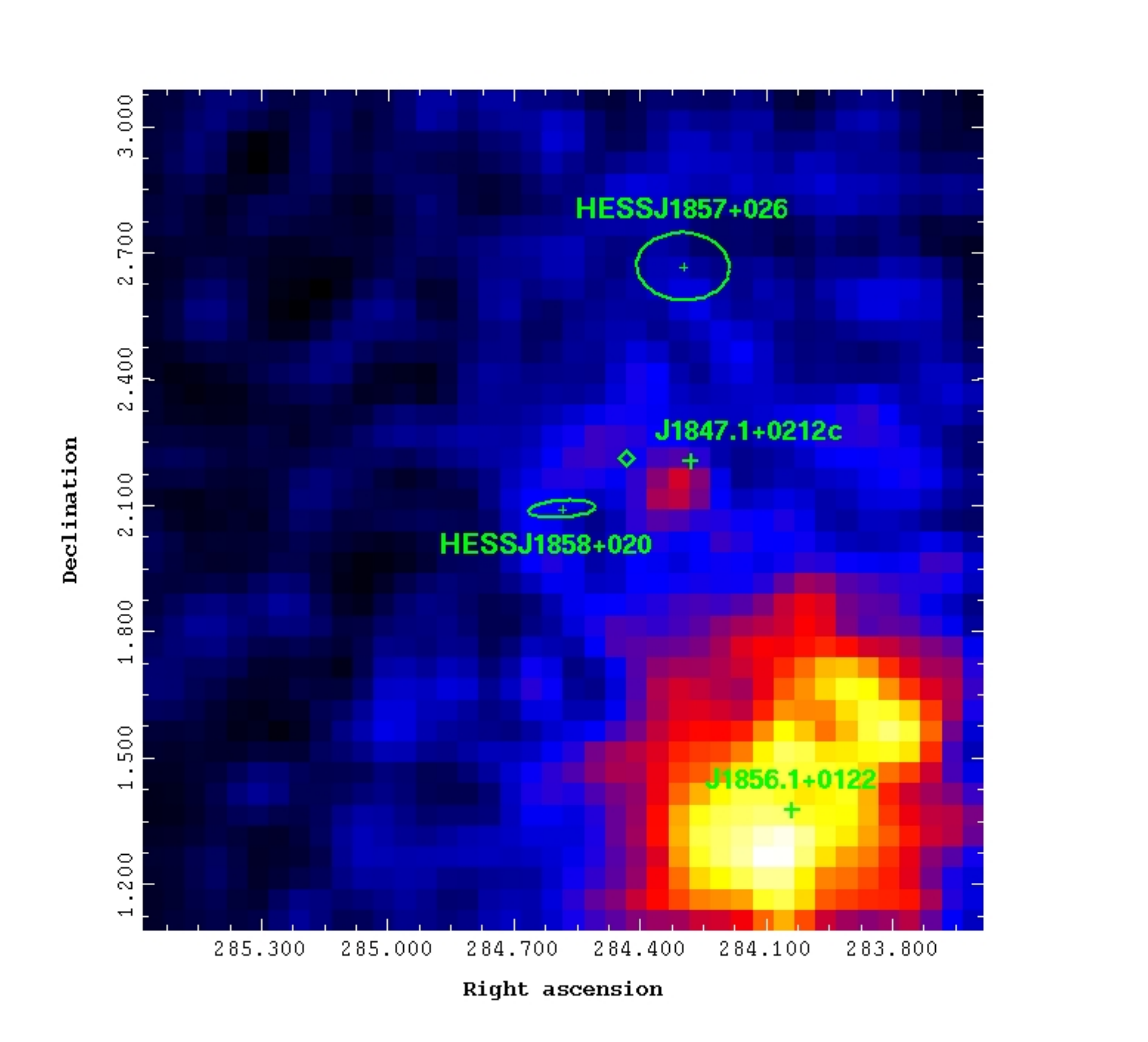}
 \caption{Counts map of Fermi-LAT data for E > 1 GeV with subtracted Galactic and isotropic diffuse
emissions, in a window of 2¼? 2¼  centered on HESS J1858+020. A gaussian smoothing with sigma=0.2¼ is applied.
The \fermi-LAT sources from the 1FGL 
catalog are marked. The center of the H.E.S.S. sources in the field are
labeled with a small cross, while their extension is represented by
the ellipsoidal fit from Aharonian et al. 2008. The position of PSR B1855+02 is marked with a diamond. 
}
\label{map}
\end{figure}

{ Figure \ref{map} shows the counts map of \fermi-LAT data for $E>1$ GeV with subtracted Galactic and
isotropic diffuse emissions, in a window of 2$^o \times 2^o$ centered on
HESS J1858+020.}

\subsection{Testing the hypothesis of an association with the closest \fermi\ source}

HESS J1858+020 is in the Galactic plane at coordinates l=35.578\dg, b=-0.581\dg\ (RA=284.584\dg, DEC=2.09\dg, { J2000}) and
does not spatially coincide with any source listed in the 1FGL. However, this region is crowded with gamma-ray sources. Indeed, in the spectral-spatial model created for the likelihood analysis we collected 93 \fermi-LAT sources, 
with 1FGL J1857.1+0212c {being} the closest source to  HESS J1858+020.
Even though the H.E.S.S. source is outside the error box of 1FGL J1857.1+0212c, an association can not be excluded a priori, because it is known that the Galactic diffuse model used to build the 1FGL catalog has the highest uncertainties on the Galactic plane, amounting to $\leq 6\%$ as estimated in the case of the supernova remnants W51C and W49 (Abdo et al. 2009, 2010c; see also \S\,4.7 of the 1FGL catalog). Furthermore, we are analyzing this region using more than 2 years of \fermi-LAT data, against the 11 months used for the 1FGL, which may induce changes in the position and fluxes of the sources in the spatial-spectral model.
 
To investigate the hypothesis of a possible association of 1FGL J1857.1+0212c with HESS J1858+020, we performed  two likelihood analyses. In one we set in the spectral-spatial model the coordinates of 1FGL J1857.1+0212c at the 1FGL position. In the other,
we set it at the position of the H.E.S.S. source. The analyses are performed setting free the flux parameters of all the point-like sources closer than 3$^{\circ}$ to HESS J1858+020, and modeling 1FGL J1857.1+0212c with a power-law with index and flux free. The only difference between both analyses was the assumed position of
1FGL J1857.1+0212c. This method  allows testing whether the position listed in the 1FGL Catalog is sustained. 
In the first case we obtained a test statistic value TS=863, while in the second case, with 1FGL J1857.1+0212c  displaced, we obtained TS=509.
The comparison of the two TS values obtained changing the coordinates of 1FGL J1857.1+0212c suggests that the association is significantly unlikely, given that $\Delta TS > 350$.  
%In terms of standard deviations, $\sigma \sim \sqrt{\Delta TS}$, the former result implies that the two models deviate  $\sim 19 \sigma$ from each other, with the more significant detection happening for a non-coincident 1FGL J1857.1+0212c.

 In order to take into account the systematics due to the uncertainties of the Galactic diffuse model, the analyses were repeated by artificially changing the normalization of the Galactic diffuse model by $\pm 6\%$ (see. e.g., Abdo et al. 2010d). Also in these cases, the differences of the TS values (for the putative change in position of 1FGL J1857.1+0212c) was large ($\Delta TS > 300$), and suggests that the association is significantly unlikely. Thus, we believe we can safely entertain the hypothesis that HESS J1858+020 and 1FGL J1857.1+0212c are not associated and proceed to impose upper limits.

\subsection{Upper limits}

\begin{table*}
\begin{center}
\label{tab1}
\caption{\fermi-LAT upper limits at the position of HESS J1858+020.}
\begin{tabular}{llll}
\hline \hline
Energy  & $F^{95\% {\rm UL}} $  & $G^{95\% {\rm UL}} $ &$\nu F_{\nu}^{95\% {\rm UL}} $ \\
 (GeV)             &  $ 10^{-9} \times$ ph cm$^{-2}$s$^{-1}$ & $ 10^{-11} \times$ \ergscm2  & $ 10^{-11} \times$ \ergscm2   \\
\hline
%bayesian
0.10 - 1.00        &  50    &  1.85    &  0.72  \\ 
1.00 - 10.00      &  6.93      &  2.56    &  1.00  \\
10.00 - 100.00  &  0.21      &  0.76    &  0.30  \\
$>$0.10                &  113  &  7.33    &    --     \\ 
%profile
%0.1--1.0  &  30  &  1  &  0.4  \\
%1.0--10  &  5.5  &  2.0  &  0.79  \\
%10--100  &  0.12  &  0.46  &  0.18  \\
%$>$0.10   &  97  &  6.3  &  --  \\
\hline
\hline
\end{tabular}
\end{center}
\end{table*}

Once the hypothesis of association is rejected, the significance of the plausible gamma-ray emission from HESS J1858+020 was evaluated by means of adding
an extra source at its position in the spectral-spatial model for the likelihood analysis. We model it with a power-law. When redoing the analysis,
we set free the photon index and flux parameters of HESS J1858+020, and the flux parameter of all the point-like sources closer than 3$^{\circ}$, with the exception of 1FGL J1857.1+0212c. As discussed above, this is  the closest 1FGL source to the position of interest; and for it, we fixed all its parameters.
The latter choice can be understood if one takes into account that the \fermi-LAT PSF at 1 GeV (around the peak of the LAT sensitivity) is 0.8$^o$. If a new source is supposed to be at a smaller distance from a bright one that is already known in the spectral-spatial model, we expect that the likelihood fitting procedure will not suppress it even if fake, but will simply share the photon counts among them, resulting in a good significance (TS$>$25) for both sources. Such case has been found, for instance,
for SGR 1627$-$41 (Abdo et al. 2010d), where it was found that the high TS derived by the {\tt gtlike}
analysis has been caused by the presence of the rather
strong unidentified source (1FGL J1636.4$-$4737), which lies
at 0.12\dg\ from the magnetar (although as in this case, it was positionally incompatible with the magnetar).

With these settings, the HESS J1858+020 likelihood analysis always results in a $TS<$25, implying no detection.
After fitting, we derived 95\% flux upper limits for $E>$100 MeV with the profile method, increasing the flux obtained for HESS J1858+020 until the maximum log likelihood decreases by 2.71/2. In the same way, the 95\% flux upper limits were evaluated in three energy bins ($0.1<E<$100 GeV) fixing the photon index to 2.26 { (the result of the likelihood fit is 2.26 $\pm$ 0.13)}. The uncertainties of the \fermi-LAT effective area, and of the Galactic diffuse emission are the two main sources of systematics that can affect the results derived with our analysis. We estimated these systematics effects by repeating the upper limits analysis using modified instrument response functions that bracket the ``Pass 6 v3 Diffuse"  effective areas, and changing the normalization of the Galactic diffuse model artificially by $\pm 6\%$. 
In addition to the evaluation of 95\% flux upper limits ($F^{95\% {\rm UL}}$),
we derived the energy flux upper limits ($G^{95\% {\rm UL}} = \int_{\Delta E} E (dN/dE) dE
$), and the SED points ($\nu F_{\nu}^{95\% {\rm UL}}$).

%bayesian
In the same way, we have also evaluated the upper limits on the base of Helene's Bayesian method (Helene 1983). The main difference respect the profile method is that in this case the 95\% upper limits are found integrating the likelihood profile (function of the source flux $F$) starting from $F=0$, without any assumption on its distribution. 
Whereas the results for the upper limits obtained with the two methods were similar, the Bayesian one gave those more conservative. 
The results of this analysis are reported in Table 1.

\subsection{Additional checks}

We additionally tested that by changing the energy threshold (from 100 to 200 MeV),
the \fermi\ Science tools (from {\tt v9r17p0} to {\tt v9r18p6}), slightly
different selection cuts in {\tt gtmktime}, and
different number of sources in the background model (including those in a ROI with size from 10$^o$ to 20$^o$),
all results are stable. 
Additionally, we have repeated the analyses using an updated version of the \fermi-LAT Catalog sources,
built using two years of data and thus more compatible with the data of the HESS J1858+020 region we focus upon. With this catalog (which is yet internal to the \fermi-LAT collaboration), the number of sources closer than 20$^o$ from the HESS J1858+020 source are 107, while those closer than 3$^o$ are 10. No significant change in the previous results where found. 
We have also considered that some of the closest sources
could be extended, and
also found the results to be stable. { We have also re-computed the upper limits with different values of the assumed photon index and found the results to be stable too. For instance, assuming a value of 2.1 for the photon index, which is the mean value of the diffuse emission spectrum, only affects the upper limit in the highest energy bin by about 1\%.}

\section{Models}

{ In this Section we follow the analysis of Aharonian \& Atoyan (1996) for a point-like injection, and of
Li \& Chen (2010)  for a non-point-like injection of cosmic-rays in order to assess the hadronic production
of gamma-rays in the proximity of SNR G35.6-0.4. See these references for notation and further clarifications. 
In addition of those cosmic-rays injected by the SNR,
we also consider the contribution of diffuse Galactic protons, as measured in the Solar neighborhood
(see, e.g., Dermer 1986).
For the gamma-ray spectrum calculations we use the analytic photon emissivity
$dN_{\gamma}/dE_{\gamma}$ developed by Kelner et al.\ (2006). }

In order to apply the non-point like approach we consider
that the  high-resolution radio image of SNR G35.6-0.4 (Green 2010) shows an extent of 15$\times$11 arcmin$^2$ and thus we approximate the average radius as $\Rs \sim 20$pc using the distance of 10.5 kpc based on the proximity of the remnant with the HII region G35.5-0.0 (Paron \& Giacani 2010). The remnant radius and age at the transition from the Sedov phase to the radiative phase are given by (Lozinskaya 1991):
\begin{eqnarray}
R_{\rm cool} &=& 20 E_{51}^{0.295} n_0^{-0.409} {\rm  pc}, \nonumber \\
t_{\rm cool}    &=& 2.7 \times  10^4 E_{51} ^{0.24} n_0^{-0.52} {\rm yr},
\end{eqnarray}
respectively. If we assume the SNR explosion energy $E_{\rm SNR}=10^{51}$erg and the ambient gas density $n_0$ of order 1 cm$^{-3}$,
G35.6-0.4 seems to be near the transition time, of age $\sim27$ kyr.
Thus we use the Sedov (1959) law  for the previous evolution of G35.6-0.4:
\be
\Rs = 0.34 \left(\frac{E_{51} }{\mu n_0} \right)^{0.2} \left(\frac{t}{\rm yr}\right)^{0.4} {\rm pc}.
\ee
We shall adopt different values for the diffusion coefficient, taking into account the  slow diffusion correction around SNRs are possible
(e.g., Fujita et al. 2009, Torres et al. 2010) and an efficiency 
$\eta=0.1$ (e.g., Blandford \& Eichler 1987), i.e., a total injection in cosmic-rays of $10^{50}$ erg (10\% of the explosion power) along the SNR lifetime. { In any case, the transition from the Sedov phase to the radiative phase 
is not sharp; and as usual, key parameters (such as distance, accurate radius, ambient
density, and explosion energy) are uncertain. Because of this, we also explore possible fits with the simple point-like injection
approach, more appropriate for the larger timescales.}

\subsection{Point-like injection}

\begin{figure*}
\centering
 \includegraphics[width=.32\textwidth, angle=0]{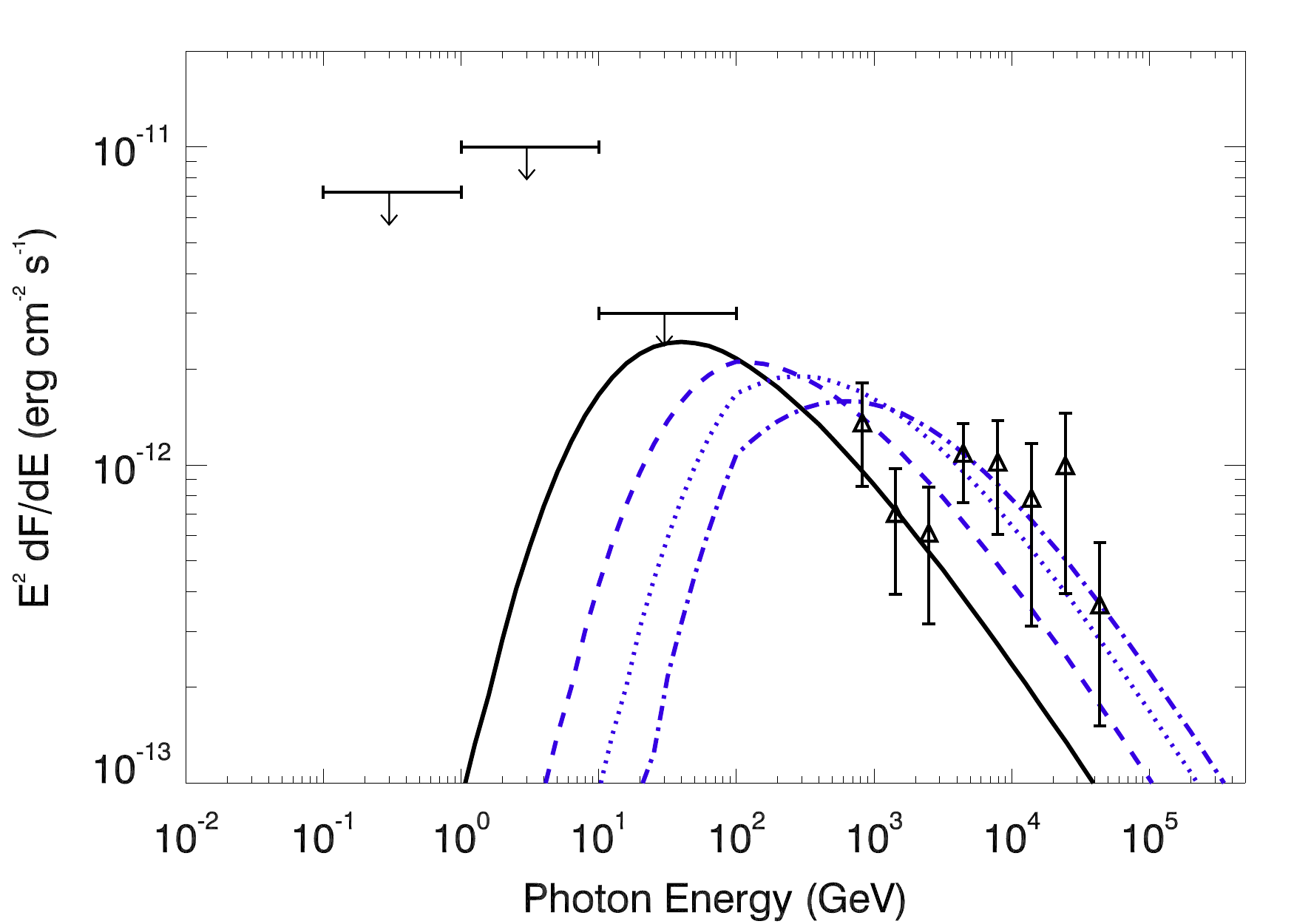}
 \includegraphics[width=.32\textwidth, angle=0]{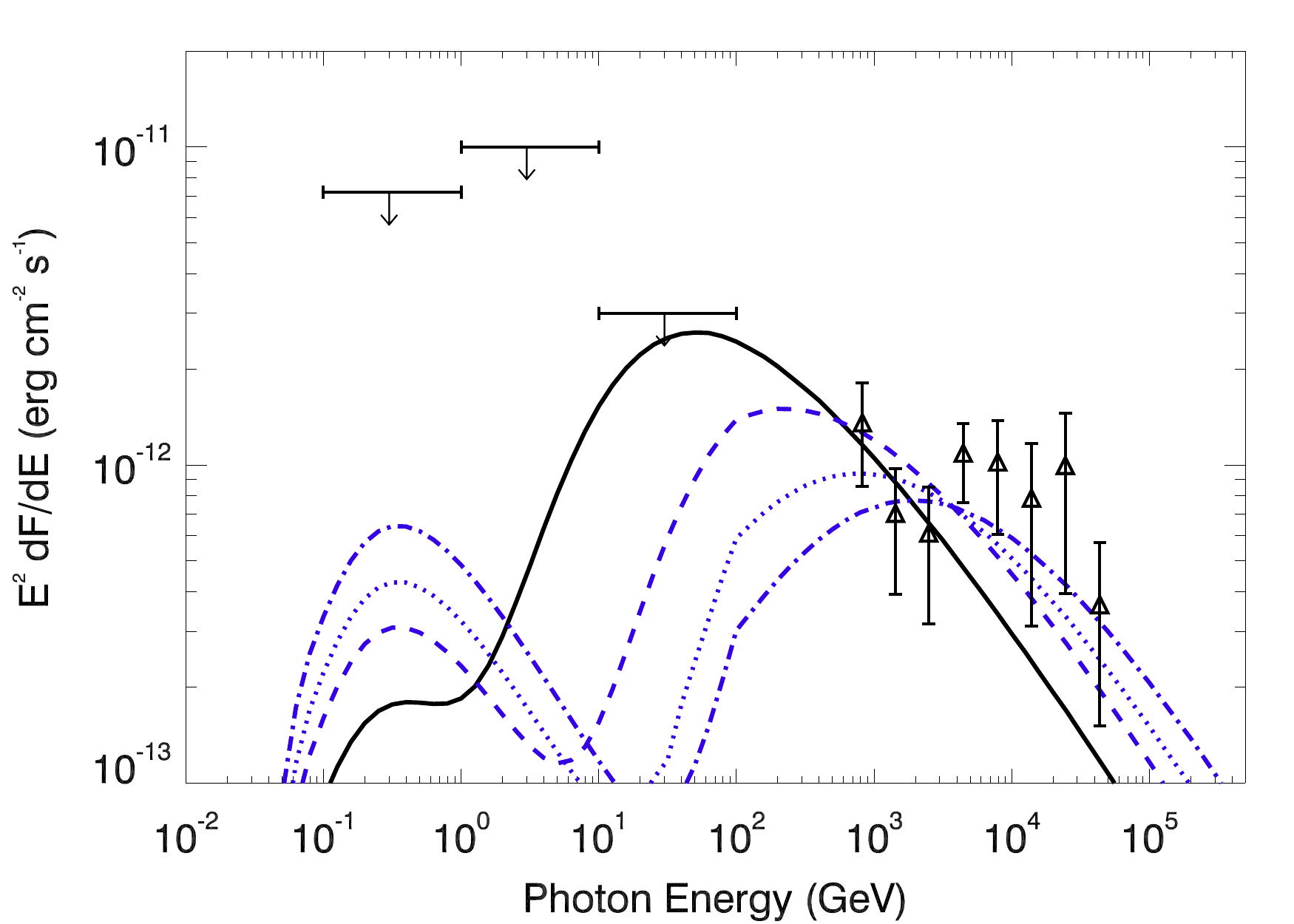}
  \includegraphics[width=.32\textwidth, angle=0]{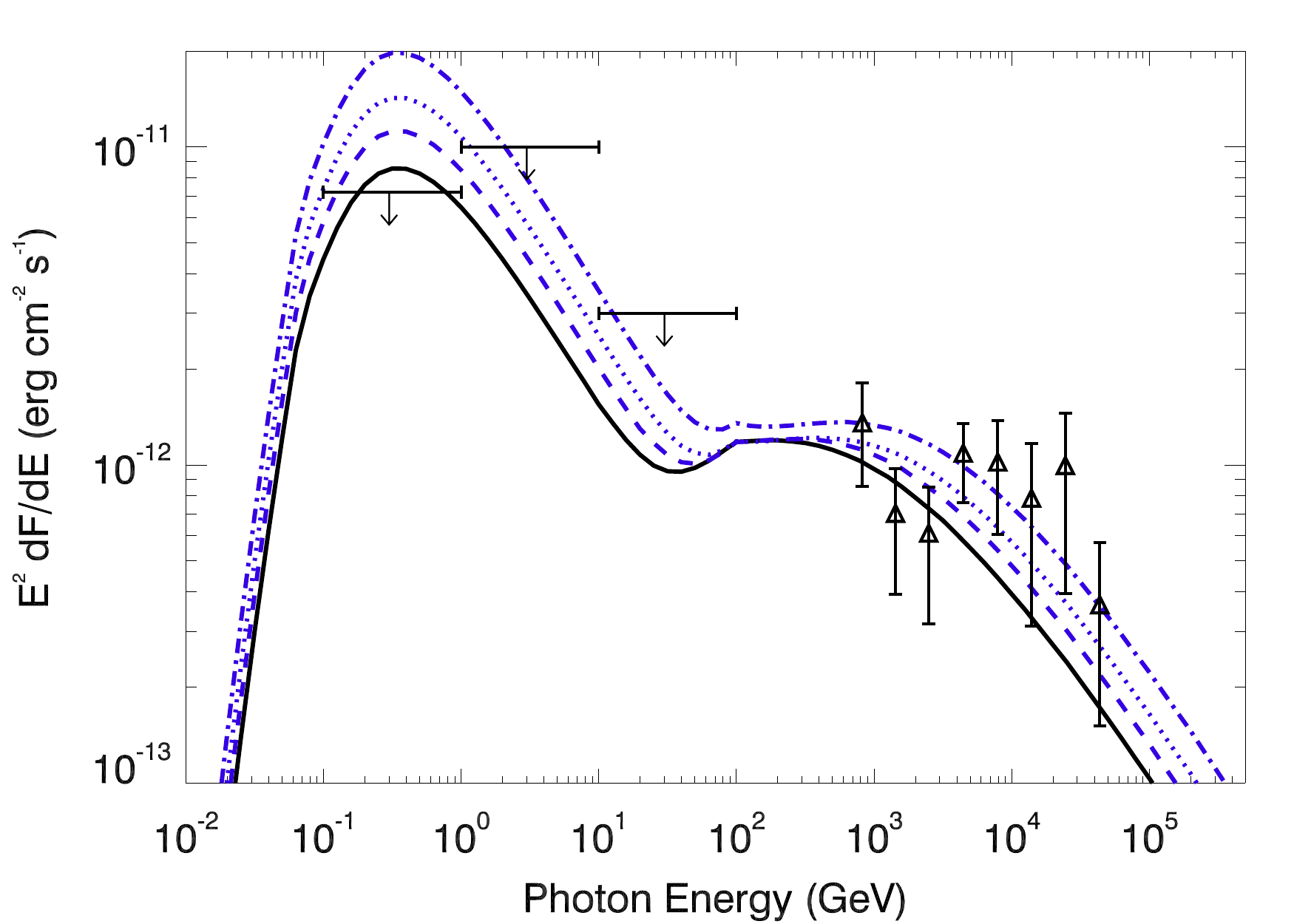}
\includegraphics[width=.32\textwidth, angle=0]{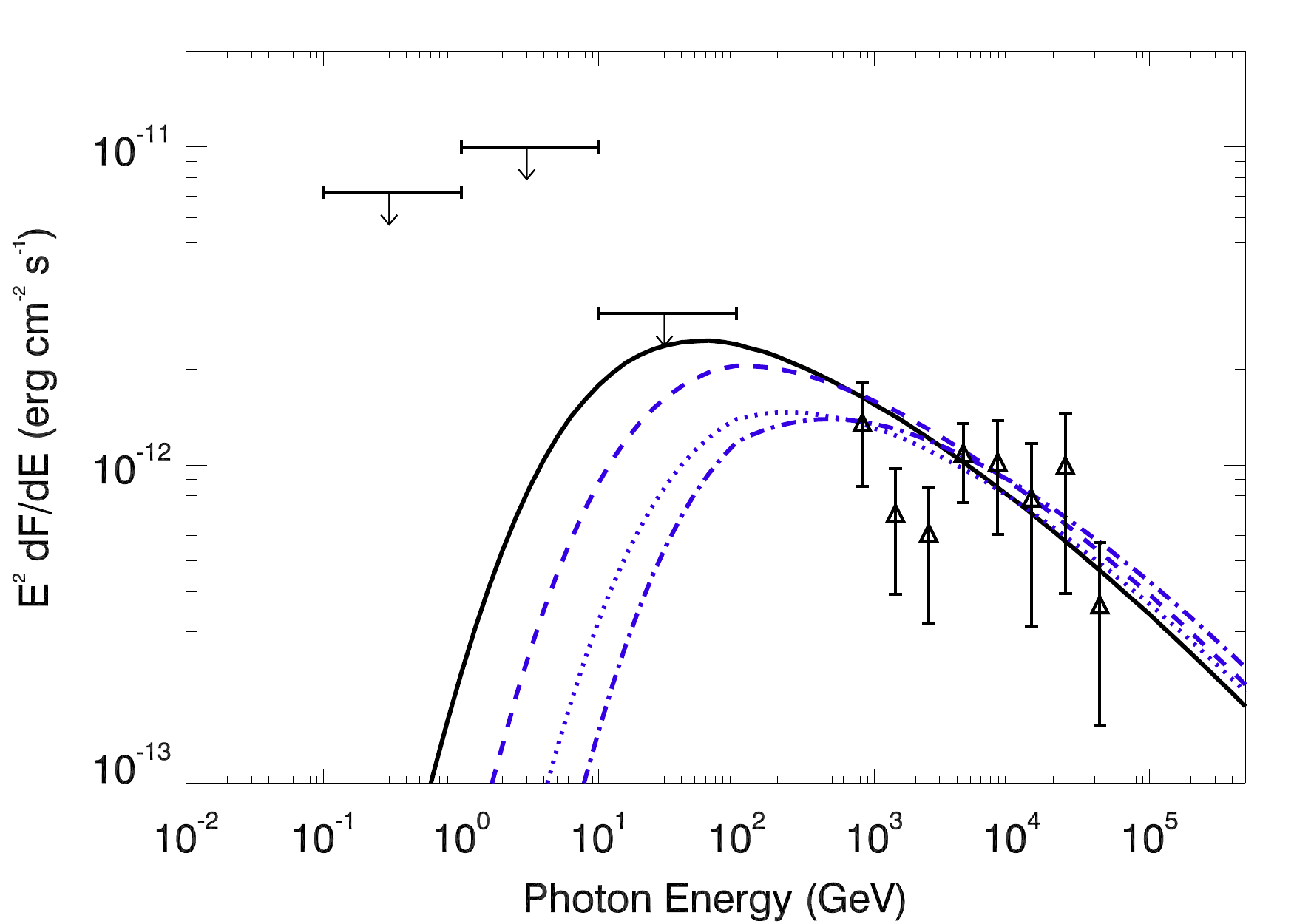}
\includegraphics[width=.32\textwidth, angle=0]{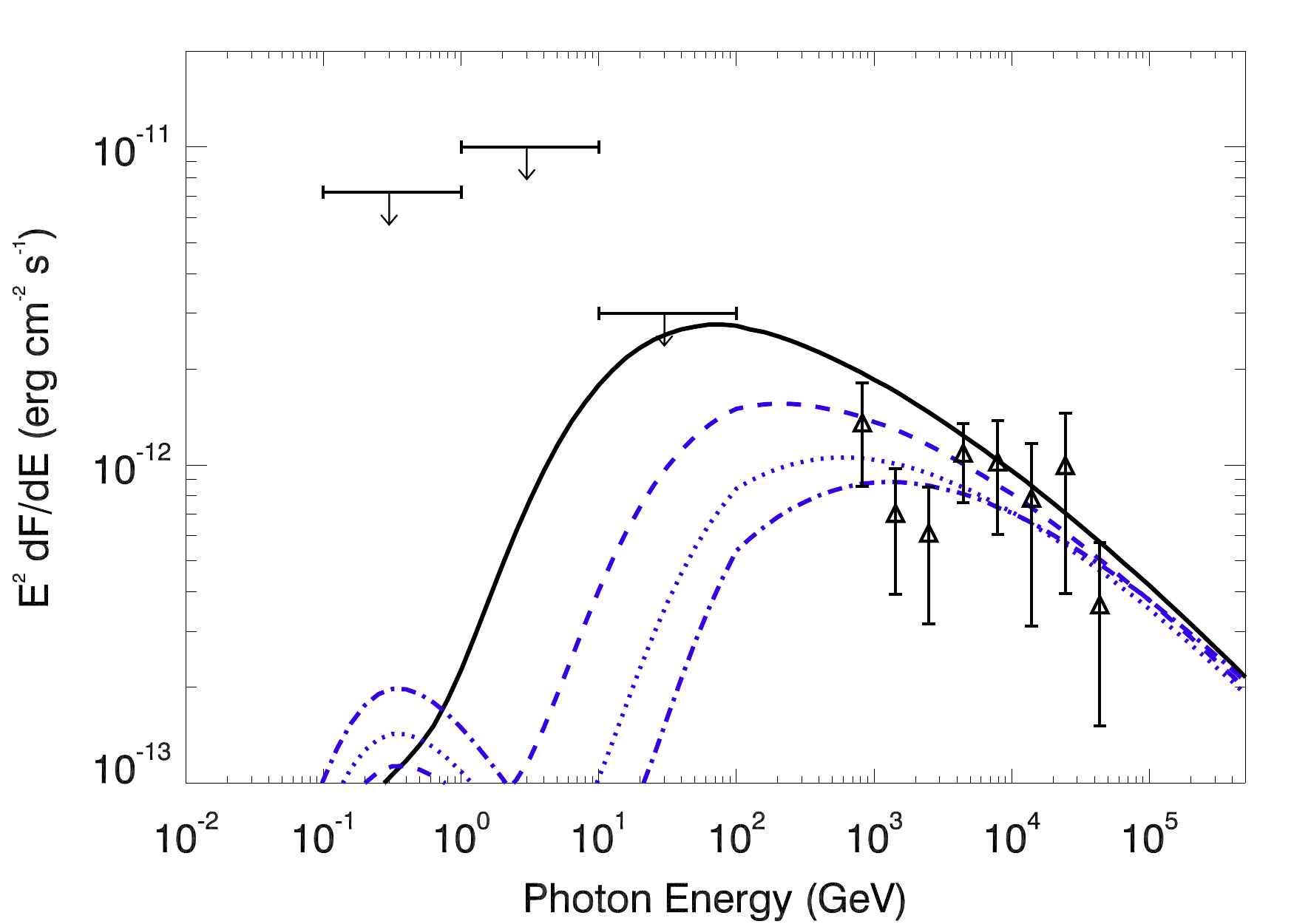}
\includegraphics[width=.32\textwidth, angle=0]{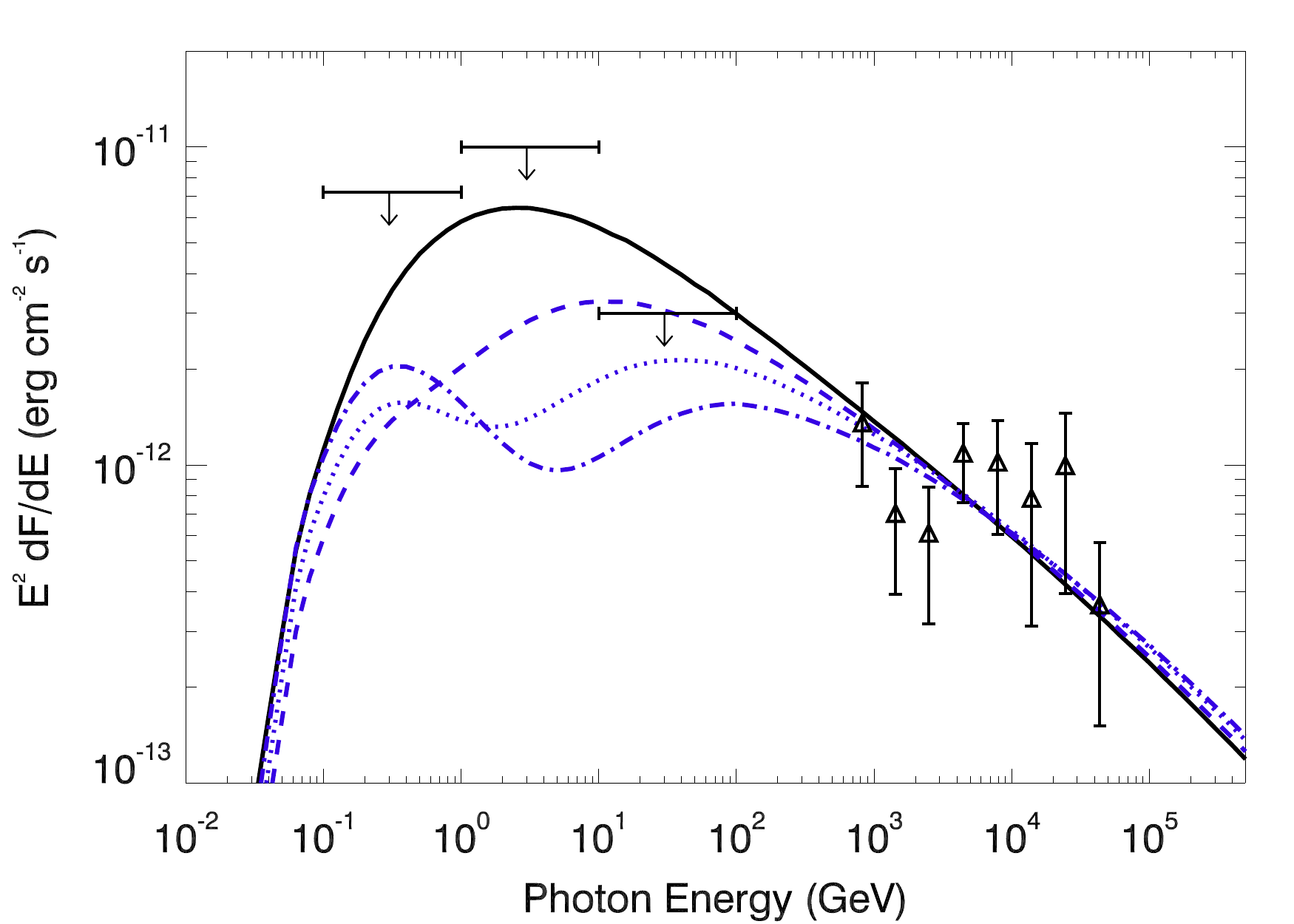}
 \caption{Point-like injection model for the gamma-ray spectrum of HESS J1858+020, compared with 
  H.E.S.S. observations (Aharonian et al. 2008) and \fermi-LAT upper limits (this work).
  The parameters we used in this figure are in correspondence with those
  given in Table 2. The top panels show cases with impulsive injection.
  The bottom
  panels show cases with continuous injection. In both panels, from left to right, we show results with increasing value of $\chi$, from  0.001 to 0.1. In each panel, solid, dashed, dotted, and dash-dotted curves stand for results with increasing values of separation between the accelerator and the cloud, according to Table 2. }
  \label{point-like-sed}
\end{figure*}

\begin{table}
\begin{center}
\caption{Parameters of the point-like diffusion model applied to HESS J1858+020, for an impulsive and a
 continuous injection of 10$^{50}$ erg along the SNR lifetime.} 
\begin{tabular}{lllllll}
\hline \hline
kind & $R_{\rm c}$ & $p$ & $\delta$ & $M_{\rm cl}$& $\chi$ \\
        &  (pc)                     &         &                &  ($10^4M_{\sun}$) &  \\
\hline
impulsive & 27 & 2 & 0.5 & 0.08  & 0.001 \\
impulsive & 29 & 2 & 0.5 & 0.16 & 0.001 \\
impulsive & 31 & 2 & 0.5 & 0.28 & 0.001 \\
impulsive & 33 & 2 & 0.5 & 0.39 & 0.001 \\
\hline
impulsive & 40 & 2 & 0.5 & 3.36  & 0.01 \\
impulsive & 50 & 2 & 0.5 & 6.05 & 0.01 \\
impulsive & 60 & 2 & 0.5 & 8.40 & 0.01 \\
impulsive & 70 & 2 & 0.5 & 12.61 & 0.01 \\
\hline
impulsive & 120 & 2 & 0.5 & 168  & 0.1 \\
impulsive & 130 & 2 & 0.5 & 220 & 0.1 \\
impulsive & 140 & 2 & 0.5 & 280 & 0.1 \\
impulsive & 150 & 2 & 0.5 & 388 & 0.1 \\
\hline
\hline
continuous & 25.5 & 2 & 0.5 & 0.04  & 0.001 \\
continuous & 26.5 & 2 & 0.5 &  0.06  & 0.001 \\
continuous & 27.5 & 2 & 0.5 & 0.07  & 0.001 \\
continuous & 28.5 & 2 & 0.5 &  1.00 & 0.001 \\
\hline
continuous & 35 & 2 & 0.5 & 1.7  & 0.01 \\
continuous & 40 & 2 & 0.5 &  2.2  & 0.01 \\
continuous & 50 & 2 & 0.5 & 2.8  & 0.01 \\
continuous & 55 & 2 & 0.5 &  3.9 & 0.01 \\
\hline
continuous & 40 & 2 & 0.5 & 12  & 0.1 \\
continuous & 50 & 2 & 0.5 &  20  & 0.1 \\
continuous & 60 & 2 & 0.5 & 30  & 0.1 \\
continuous & 70 & 2 & 0.5 &  40 & 0.1 \\
\hline
\hline
\end{tabular}
\end{center}
\label{point-table}
\end{table}

\begin{figure*}
\centering
\includegraphics[width=.32\textwidth, angle=0]{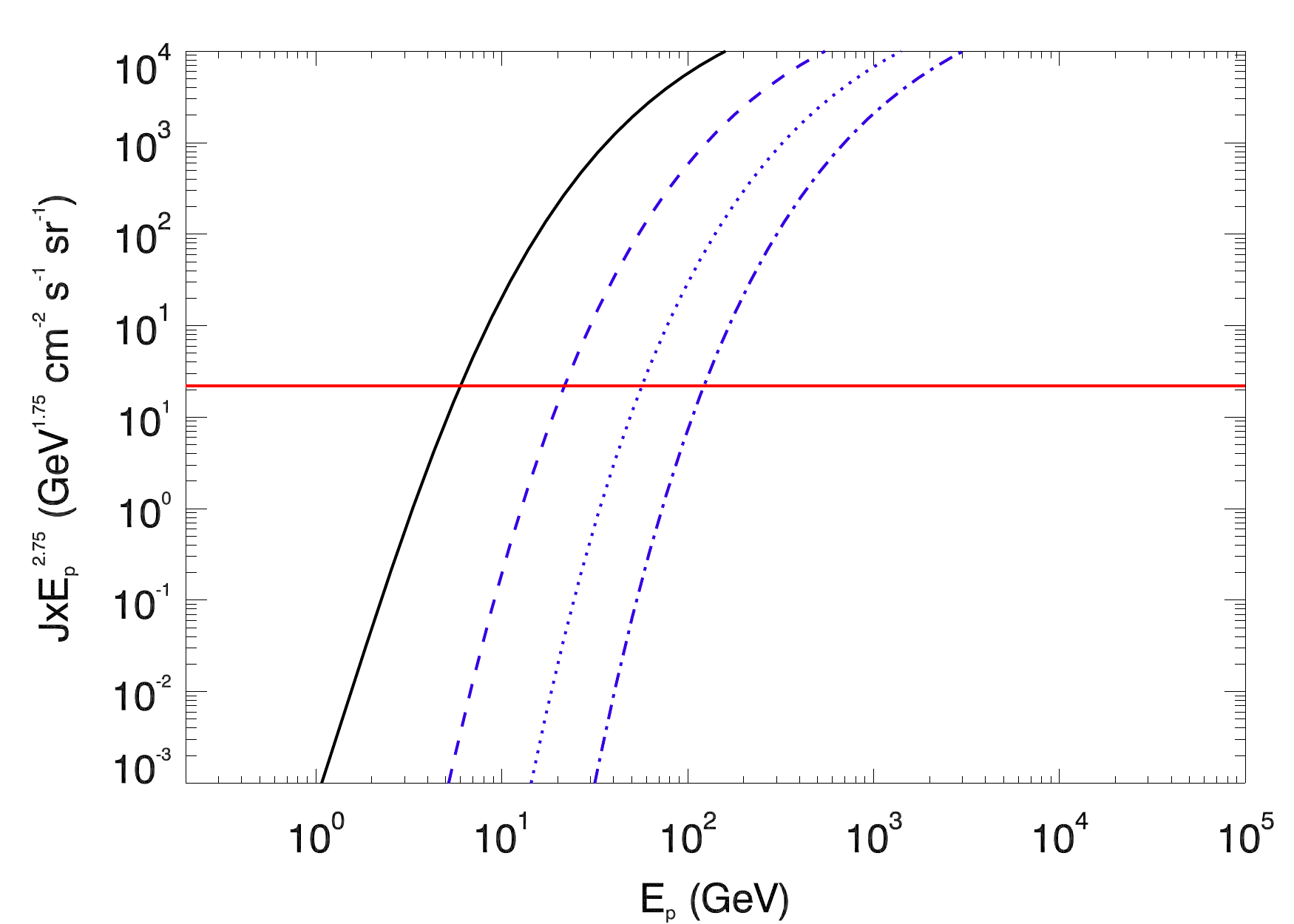}
 \includegraphics[width=.32\textwidth, angle=0]{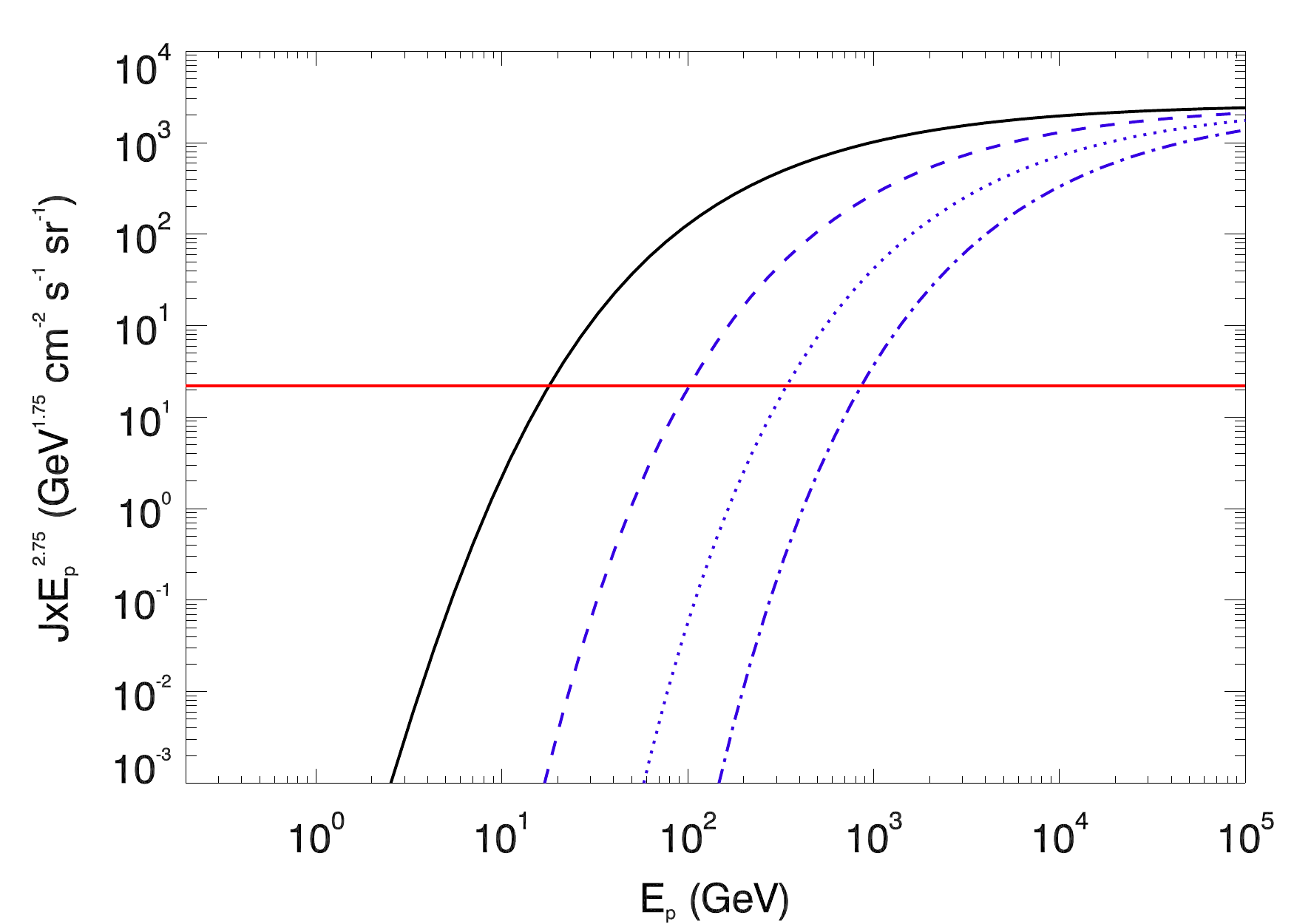}
 \includegraphics[width=.32\textwidth, angle=0]{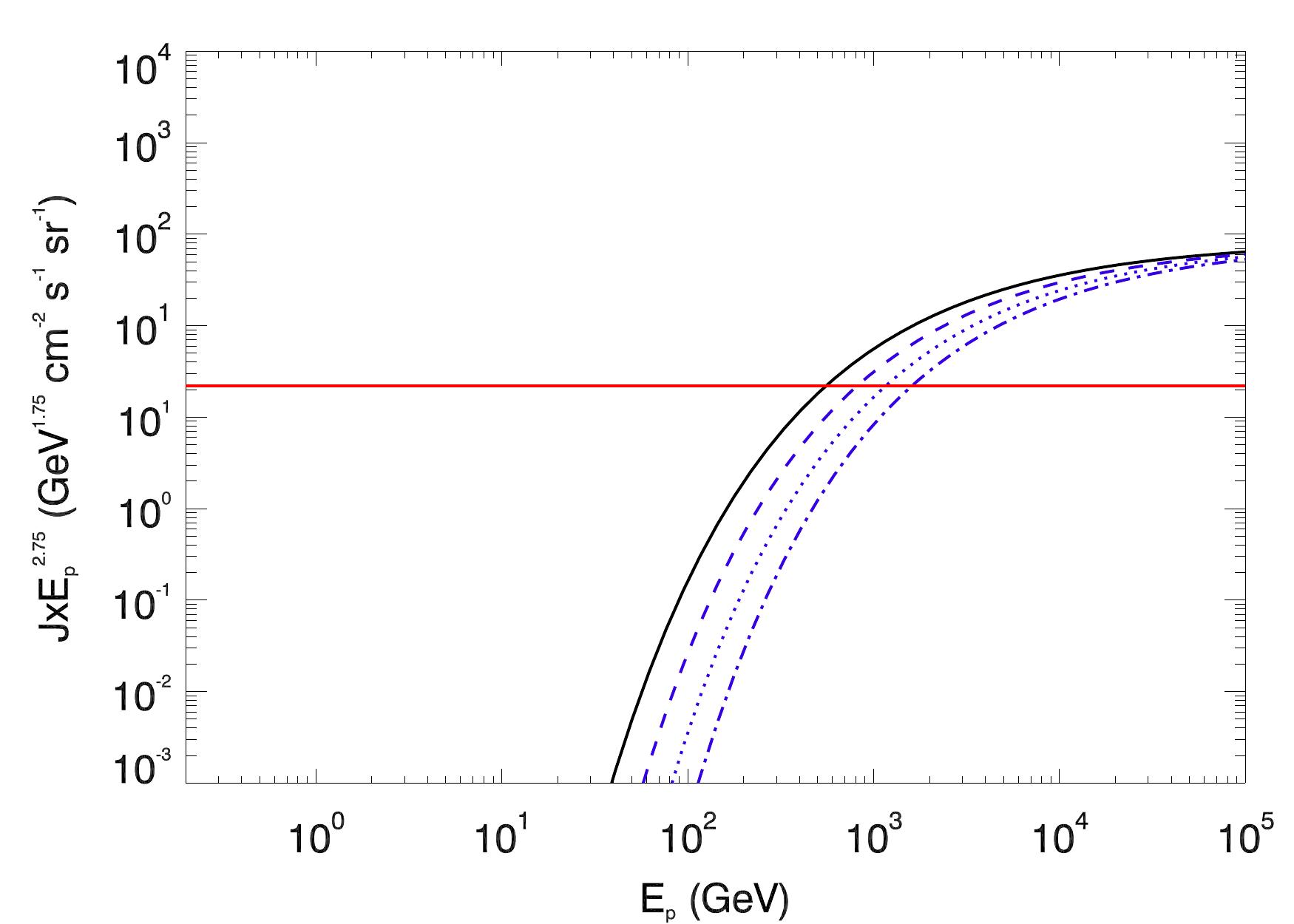}
 \includegraphics[width=.32\textwidth, angle=0]{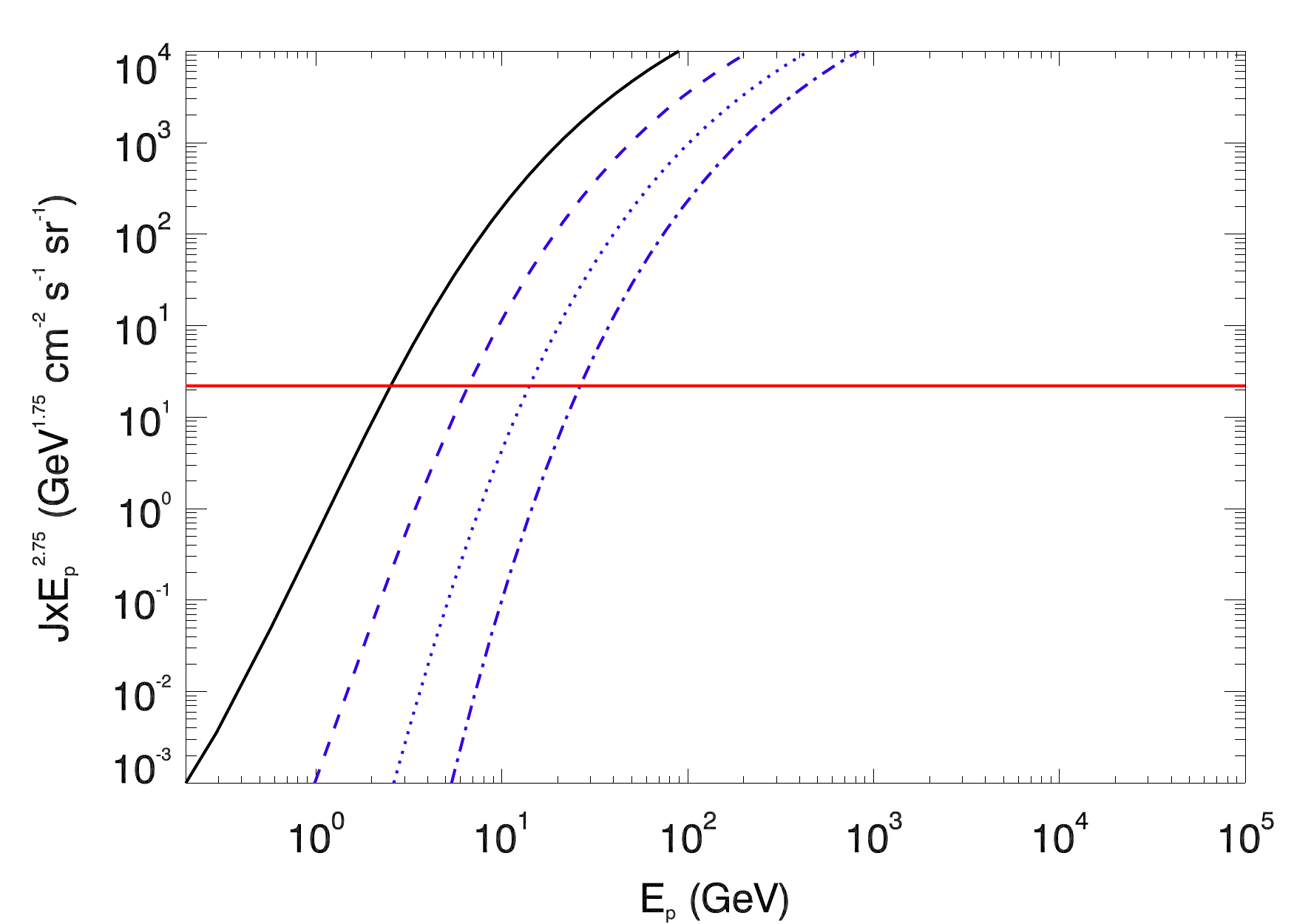}
 \includegraphics[width=.32\textwidth, angle=0]{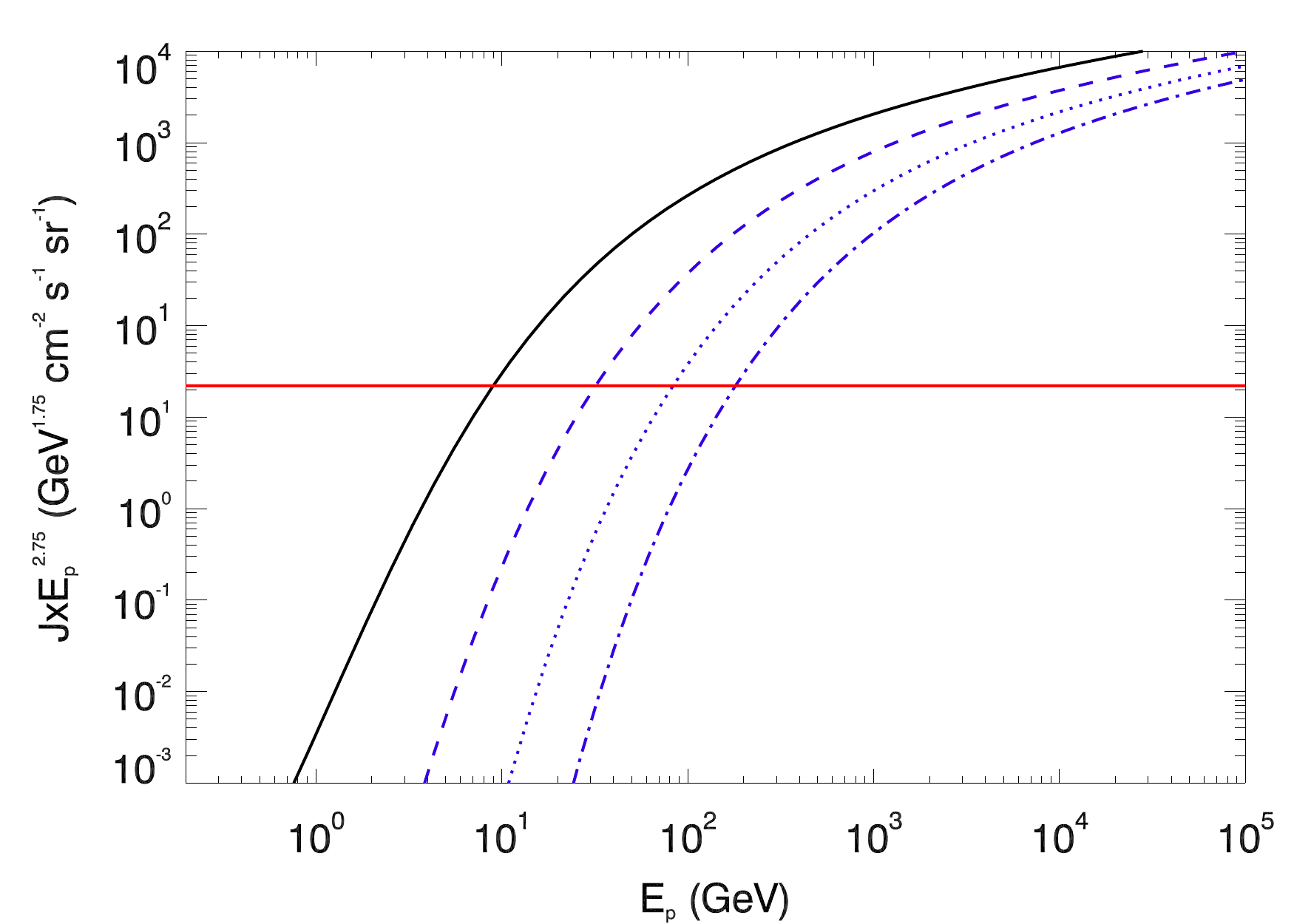}
 \includegraphics[width=.32\textwidth, angle=0]{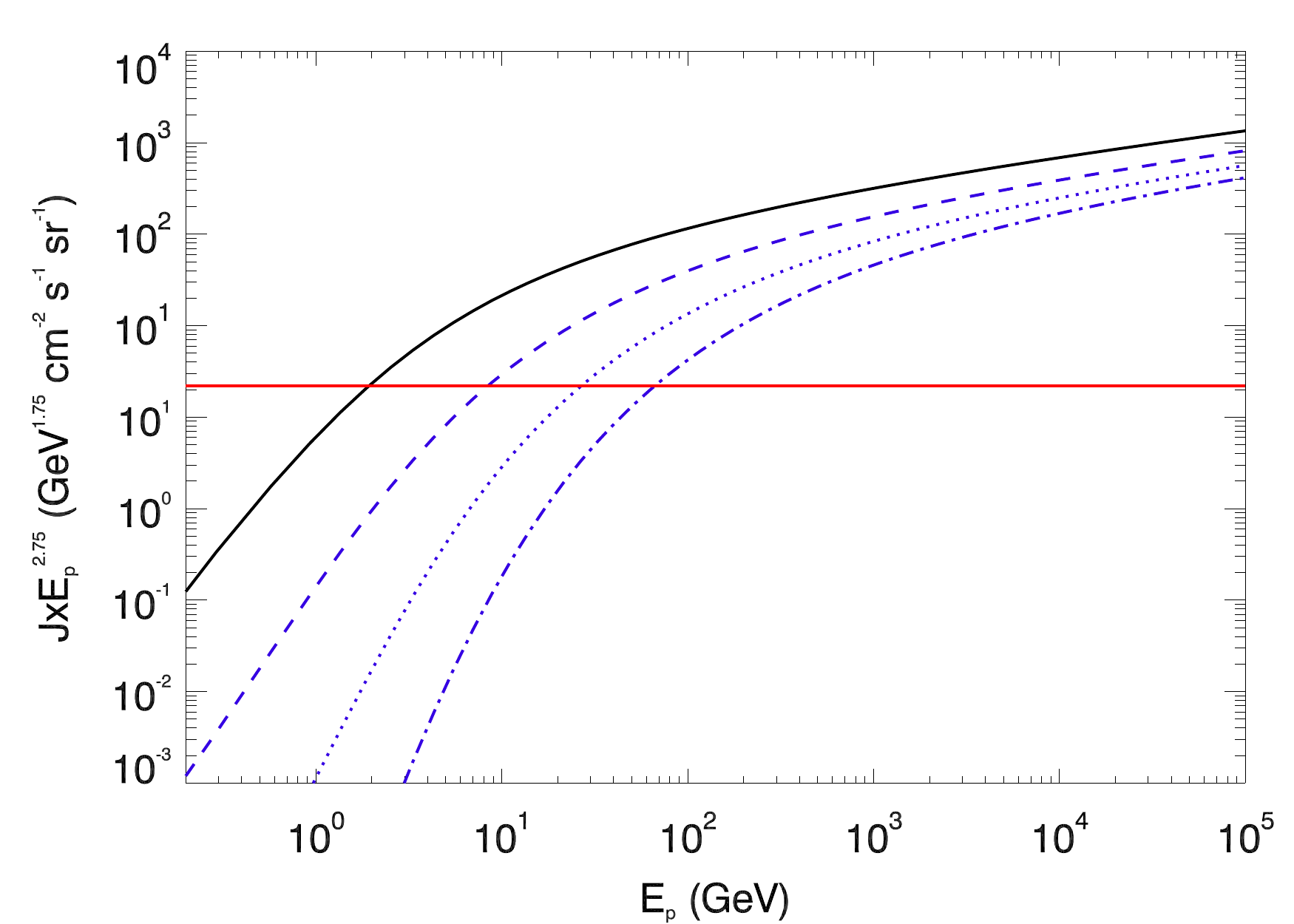}
 \caption{ Top Panel: Examples of the cosmic-ray spectrum at the position of HESS J1858+020 resulting from the point-like
 models of Figure 1 as compared with the cosmic-ray sea
{
(red horizontal line). 
}
 From left to right, 
 we plot the cases with $\chi =0.001, 0.01$ and 0.1 presented in Figure 1 and Table 2. 
 Bottom panel: {\it ibid.} but for continuous injection.}
  \label{point-like-cr}
\end{figure*}

We first adopt the point-like injection approximation and analyse which are the possible combinations of parameters, if any,
which would show the GeV--TeV phenomenology seen. Figure \ref{point-like-sed} shows some examples of these cases.
In the first place, we show the resulting SED in the gamma-ray domain for 
a point-like impulsive cosmic-ray injection, of 10$^{50}$ erg. The accelerator is placed at the center of the SNR G35.6-0.4, and the SED is the result of the interaction of the diffusing cosmic rays with the molecular material, assumed to be located in a cloud at a separate position. Table 2 and the top panels of Figure \ref{point-like-sed}
summarize the results. 
{ The diffusion
coefficient is assumed to be in the form of a power-law in energy (see, e.g., Aharonian \& Atoyan 1996)
$ D(E_{\rm
p})=10^{28}\chi(E_{\rm p}/10\rm GeV)^{\delta}\cm^2\s^{-1} , $ where
$\chi$ is the correction factor for slow diffusion around the SNR
(e.g., Fujita \etal\, 2009) and $\delta\approx0.3$--0.7 (e.g., Berezinskii
1990). The value $10^{28}\chi\cm^2\s^{-1}$ is referred to as $D_{10}$ below.}
To simplify, we have adopted a fiducial value { for the slope of the injected cosmic-ray spectrum} $p=2$,  
and $\delta=0.5$, since other values within the corresponding typical phase space of these parameters 
do not change our conclusions. Instead, we show results for different values of $D_{10}$, from $10^{25}$ to 10$^{27}$
cm$^2$ s$^{-1}$. We see that the higher the value of $D_{10}$ the more important becomes the contribution of the cosmic-ray sea with respect to that of the accelerator.
This happens because in order to have a reasonably good fit to the GeV--TeV data, { we need larger values of SNR-MC separation and MC masses.} The latter cannot attain any needed value, of course, but it is constrained to be less than the measured one in the region, which is of the order of $10^4$ M$_\odot$ (Paron \& Giacani 2010). In addition, the separation cannot be arbitrarily large, since, unless the clouds are  significantly in the foreground or background of the accelerator --which admittedly may well be the case-- the projected distance would be larger than measured. The projected distance of the MC to the SNR center is $\sim30$ pc.
These constraints exclude most models { with an impulsive injector and}  $D_{10} > 10^{26}$ 
cm$^2$ s$^{-1}$. A value of $D_{10} = 10^{27}$
cm$^2$ s$^{-1}$ with an impulsive injection from the SNR is untenable, for instance, since it would require two orders of magnitude more molecular material than available. { For the largest $D$ (top panel, right), even considering unrealistic masses and distance, the model fits for large separations are above the upper limits and seem inadequate.}
Reasonably good fits, i.e., resulting in separation and MC masses in agreement with measurements, and within the assumptions made, would require 
a very low value of the diffusion coefficient. Such values of $D_{10}$, representing a very slow diffusion, have 
been used for instance by Gabici et al. (2007), but it is likely that they require a much denser cloud
to be feasible.

In Figure \ref{point-like-sed} and Table 2
we also show several examples of the resulting SED for a continuous injection of 1.17 $\times 10^{38}$ erg s$^{-1}$ (totalizing 
10$^{50}$ erg in the lifetime of the SNR) with the accelerator also located at the SNR center. Again, reasonably good fits can only be obtained with low values of 
$D_{10}$, up to $\sim$10$^{26}$
cm$^2$ s$^{-1}$. Values of the needed molecular mass in interaction with the cosmic-ray population 
increases with $D_{10}$, but in this case, values up to $D_{10}=10^{26}$
cm$^2$ s$^{-1}$ are able to produce a good fit, whereas higher ones would require masses 
beyond those measured. 
 
In Figure  \ref{point-like-cr} we show 
the cosmic-ray spectrum at the position of HESS J1858+020 resulting from the point-like
models of Figure 1, as compared with the cosmic-ray sea. 
From left to right, 
we plot the cases with $\chi =0.001, 0.01$ and 0.1, using the same color coding and parameters as those presented in Figure 1 and Table 2. The only solutions that may be in agreement with the total molecular mass measured in this environment correspond to the leftmost top panel, and the two leftmost bottom panels in this figure. The particles injected by the accelerator dominate the cosmic-ray sea for energies above $\sim$ 10 GeV in these cases, and quickly overcome it by several orders of magnitude for higher energies.
  
We note however that the caveat of this analysis might be in its point-like assumption. The size of the SNR shell, and position of the cloud with respect to the center of the SNR are comparable, and it is thus an approximation to consider this setup. In order to explore further this environment, especially for the continuous injection,
we apply the cumulative diffusion model to explore the range of parameters with which the source HESS J1858+020
could be associated with the newly identified SNR G35.6-0.4 leaving a non-detection at GeV energies when the SNR is no longer assumed to be point-like. 
 
\subsection{Non point-like injection}

The way in which the cosmic-ray spectrum is constructed in the cumulative model is different from the more direct point-like approach. In the cumulative model, we are adding contributions coming from different distances, and thus diffusing differently. This is especially important when the MC is indeed close to the SNR shell. In this case the cosmic rays entering the cloud arrive from distances that span from a few to tens of pc (for those coming from the other side of the shell).

\begin{figure*}
\centering
 \includegraphics[width=.37\textwidth, angle=0]{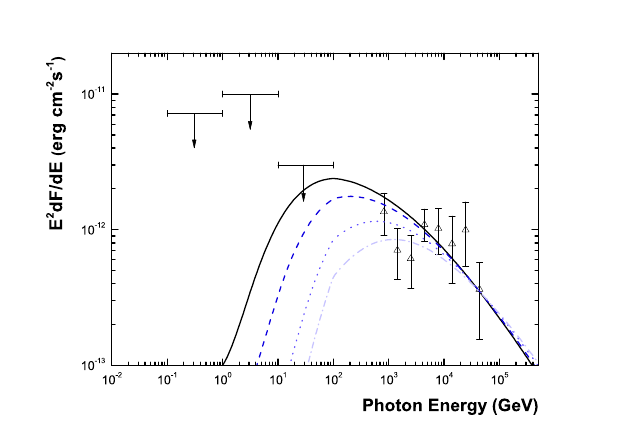}
  \hspace{-1.2cm}
  \includegraphics[width=.37\textwidth, angle=0]{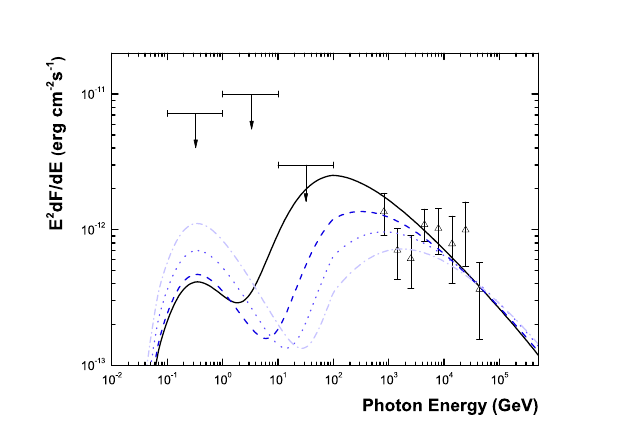}
   \hspace{-1.2cm}
   \includegraphics[width=.37\textwidth, angle=0]{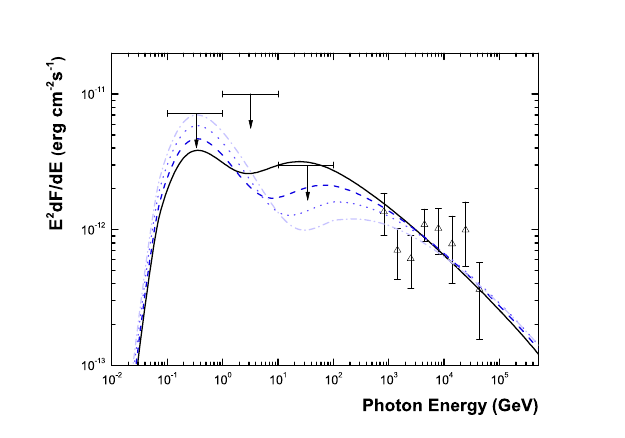}
 \caption{Non point-like injection model for the gamma-ray spectrum of HESS J1858+020, compared with 
  H.E.S.S. observations (Aharonian et al. 2008) and \fermi-LAT upper limits  (this work). From left to right, we show results with increasing value of $\chi$, from  0.001 to 0.1.
   The parameters we used in this figure are in correspondence with those
  given in Table 3. Solid, dashed, dotted, and dash-dotted curves stand for results with increasing values of separation.
  Bottom panels: }
  \label{fig:spectra}
\end{figure*}

With these SNR parameters, the GeV--TeV data of HESS J1858+020 are fitted (as plotted
in Figure \ref{fig:spectra}) and the resulting parameters are listed in Table 3.
To compare with the former case, we have adopted an initial spectral index of escaping protons, $p=2$, predicted
by the classic Fermi-type acceleration process, and $\delta = 0.5$. 
We have also explored steeper  values of $p$ to fit the gamma-ray spectrum of this source so as to
study the parameter dependence. For $p$-values equal to 2.1 or 2.2, the needed $\delta$ would
decrease a little in the range 0.3--0.5 to better fit the spectral data,
while the lower limit
of $R_{\rm c}$ would still be at $\sim$40--60 pc. 
Higher values of $p$, e.g., 2.4 and beyond, would
produce $\delta<0.3$, which is unreasonable for the cosmic-ray diffusion process.

\begin{table}
\begin{center}
\label{tab13}
\caption{Parameters of the non point-like (3D) diffusion model applied to HESS J1858+020, for an impulsive and a
 continuous injection of 10$^{50}$ erg along the SNR lifetime.} 
\begin{tabular}{cccccc}
\hline \hline
kind & $R_{\rm c}$ & $p$ & $\delta$ & $M_{\rm cl}$& $\chi$ \\
        &  (pc)                     &         &                &  ($10^4M_{\sun}$) &  \\
\hline
3D & 22 &  2  &  0.5  &  1.1 & 0.001 \\
3D & 23 &  2  &  0.5  &  1.1 & 0.001 \\
3D & 24 &  2  &  0.5  &  1.2 & 0.001 \\
3D & 25 &  2  &  0.5  &  1.6 & 0.001 \\
\hline
3D & 30 &  2  &  0.5  &  6.7  & 0.01\\
3D & 35 &  2  &  0.5  &  8.2  & 0.01 \\
3D & 40 &  2  &  0.5  &  12 & 0.01 \\
3D & 45 &  2  &  0.5  &  19  & 0.01\\
\hline
3D & 50 &  2  &  0.5  &  70 & 0.1 \\
3D & 60 &  2  &  0.5  &  80 & 0.1 \\
3D & 70 &  2  &  0.5  &  100  & 0.1 \\
3D & 80 &  2  &  0.5  &  120 & 0.1 \\
\hline
\hline
\end{tabular}
\end{center}
\end{table}

\begin{figure*}
\centering
 \includegraphics[width=.37\textwidth, angle=0]{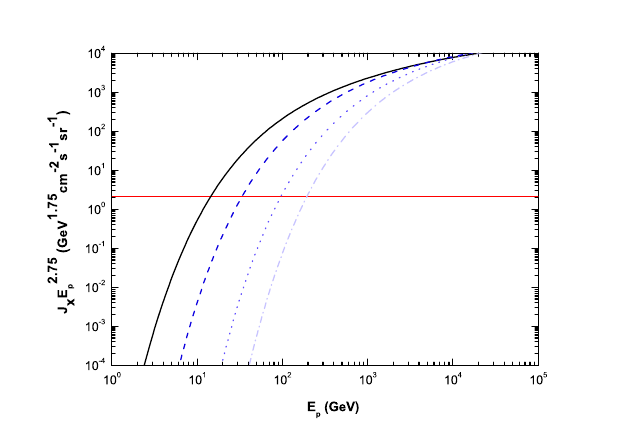}
 \hspace{-1.2cm}
 \includegraphics[width=.37\textwidth, angle=0]{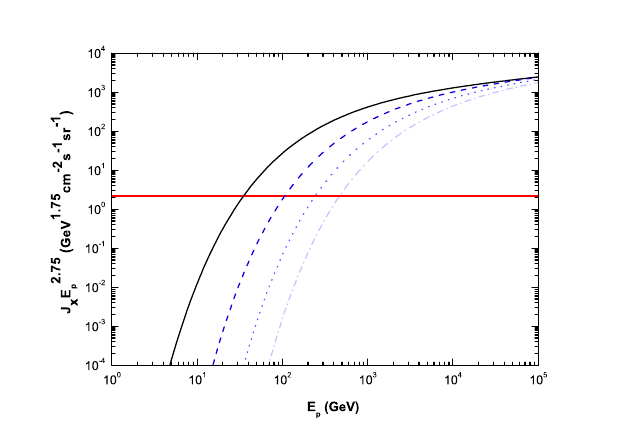}
 \hspace{-1.2cm}
  \includegraphics[width=.37\textwidth, angle=0]{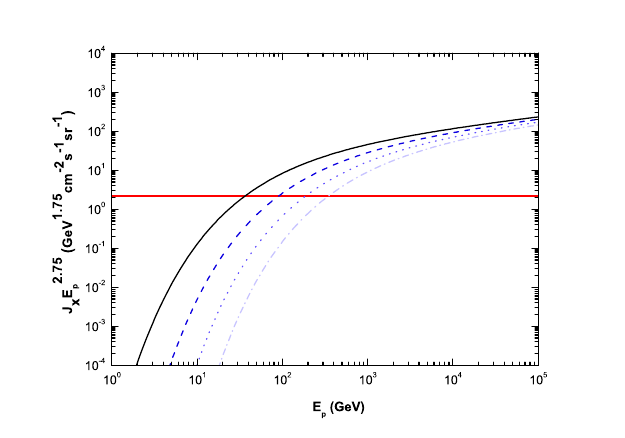}
 \caption{ Examples of the cosmic-ray spectrum at the position of HESS J1858+020 resulting from the non point-like
 models of Figure \ref{fig:spectra} as compared with the cosmic-ray sea. From left to right, 
 we plot the cases with $\chi =0.001, 0.01$ and 0.1 presented in Figure \ref{fig:spectra} and Table 3.}
   \label{point-like-noncr}
\end{figure*}

The fitting results for $\chi=0.001, 0.01,$ and 0.1 are shown in Figure 3. As expected, among the fitted parameters, the MC mass ($M_{\rm cl}$) and the distance from the SNR center to the MC ($R_{\rm cl}$) decrease when the $\chi$ value is reduced. In the higher end for the diffusion coefficient ($\chi=0.1$), the fitted $R_{\rm cl}$ value should be in the range of 50 pc to 80 pc. But also for the 3D model it can be noted that the MC mass required for $\chi=0.1$ is of order $10^6$ M$_{\sun}$, which significantly conflicts with the observed value. In the more realistic 3D model, also the case with $D_{10} = 10^{26}$
cm$^2$ s$^{-1}$ requires a mass beyond the one measured, by a factor of $\sim 6$ or more if the clouds are more distant. On the other hand, in the lower end of $D_{10}$ 
($\chi=0.001$), the distance from the SNR center to the MC is quite similar to the SNR radius itself ($\sim20$ pc), implying that the MC is essentially at the end of the shock front, which actually could be argued from the projected  maps of Paron \& Giacani (2010). Thus we find that in order for a hadronic origin of the gamma-ray emission seen be viable, 
the diffusion coefficient near this SNR should be greatly suppressed with respect to the average Galactic value, by more than one order of magnitude.

Figure \ref{point-like-noncr} shows the results for the cosmic-ray spectrum. Note especially the differences at the lowest value of $\chi$, the leftmost panel, when compared with the point-like injection. The 3D analyses more realistically describes the obtained data at GeV and TeV energies, with a much lower enhancement of cosmic rays in the region; about a factor of 100 beyond the cosmic-ray sea level at $\sim$ 1 TeV and a crossing at $\sim$ 10--100 GeV.

\subsubsection{An energy-independent diffusion?}

\begin{figure*}
\centering
 \includegraphics[width=.37\textwidth, angle=0]{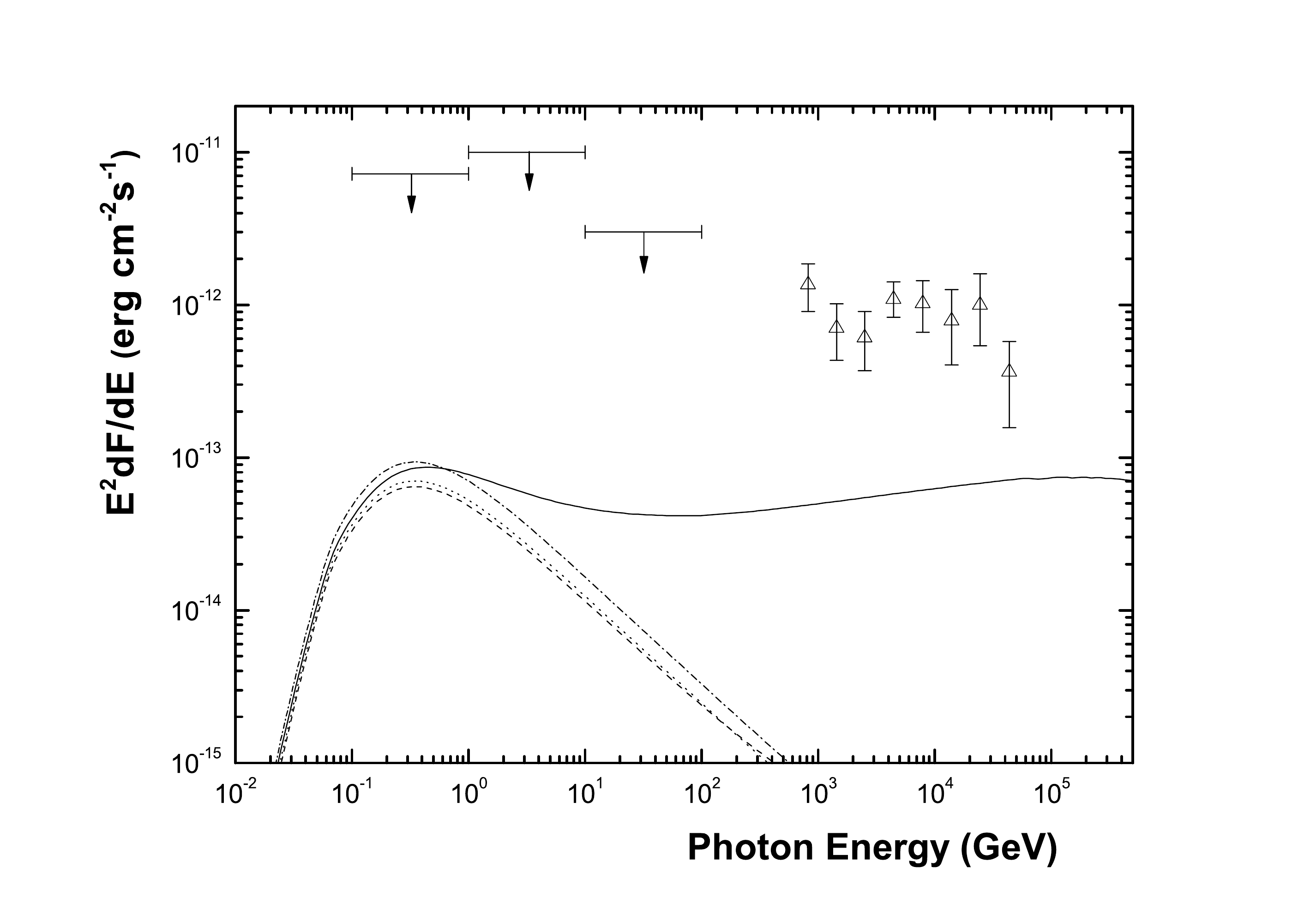}
 \hspace{-1.2cm}
 \includegraphics[width=.37\textwidth, angle=0]{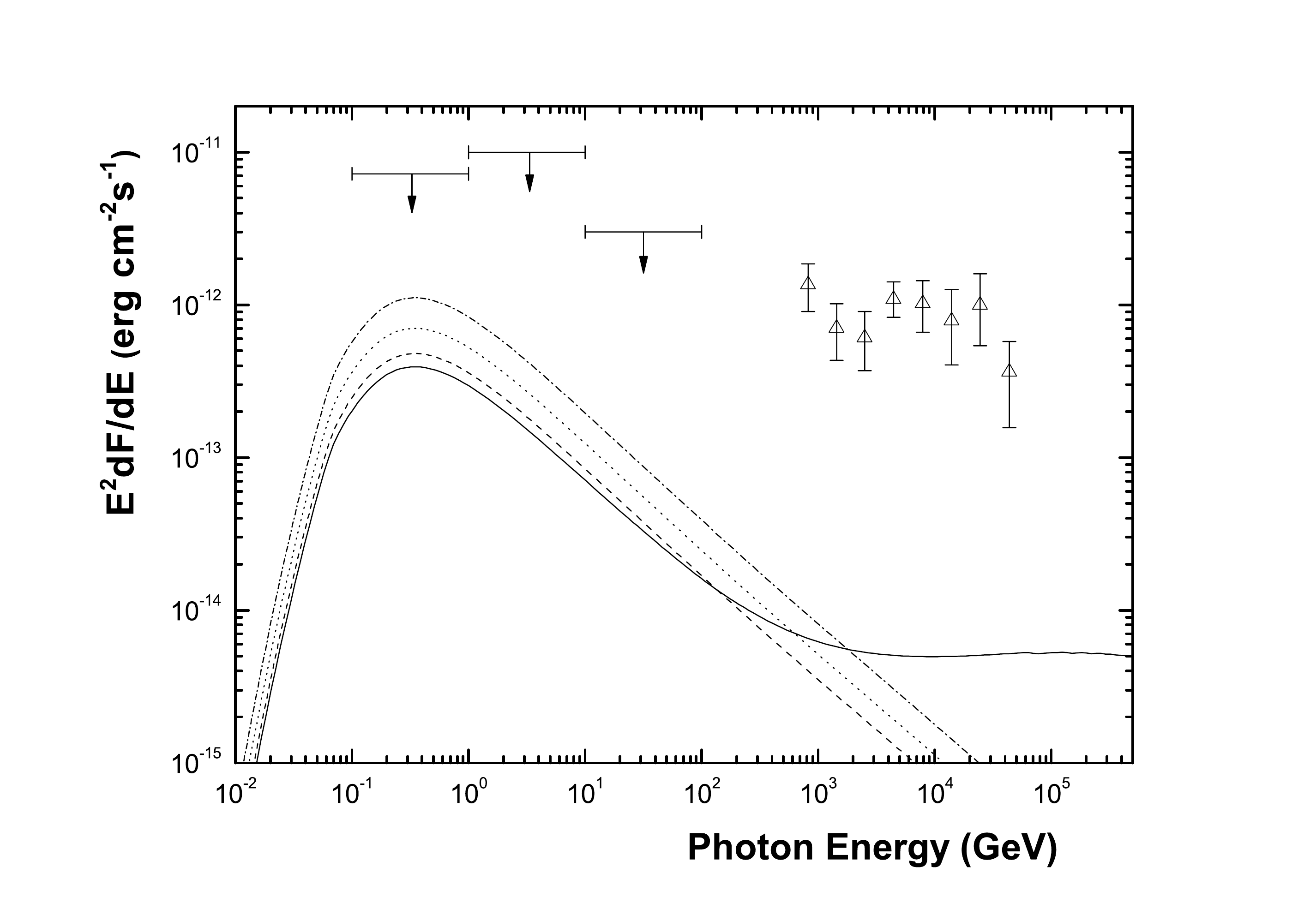}
  \hspace{-1.2cm}
 \includegraphics[width=.37\textwidth, angle=0]{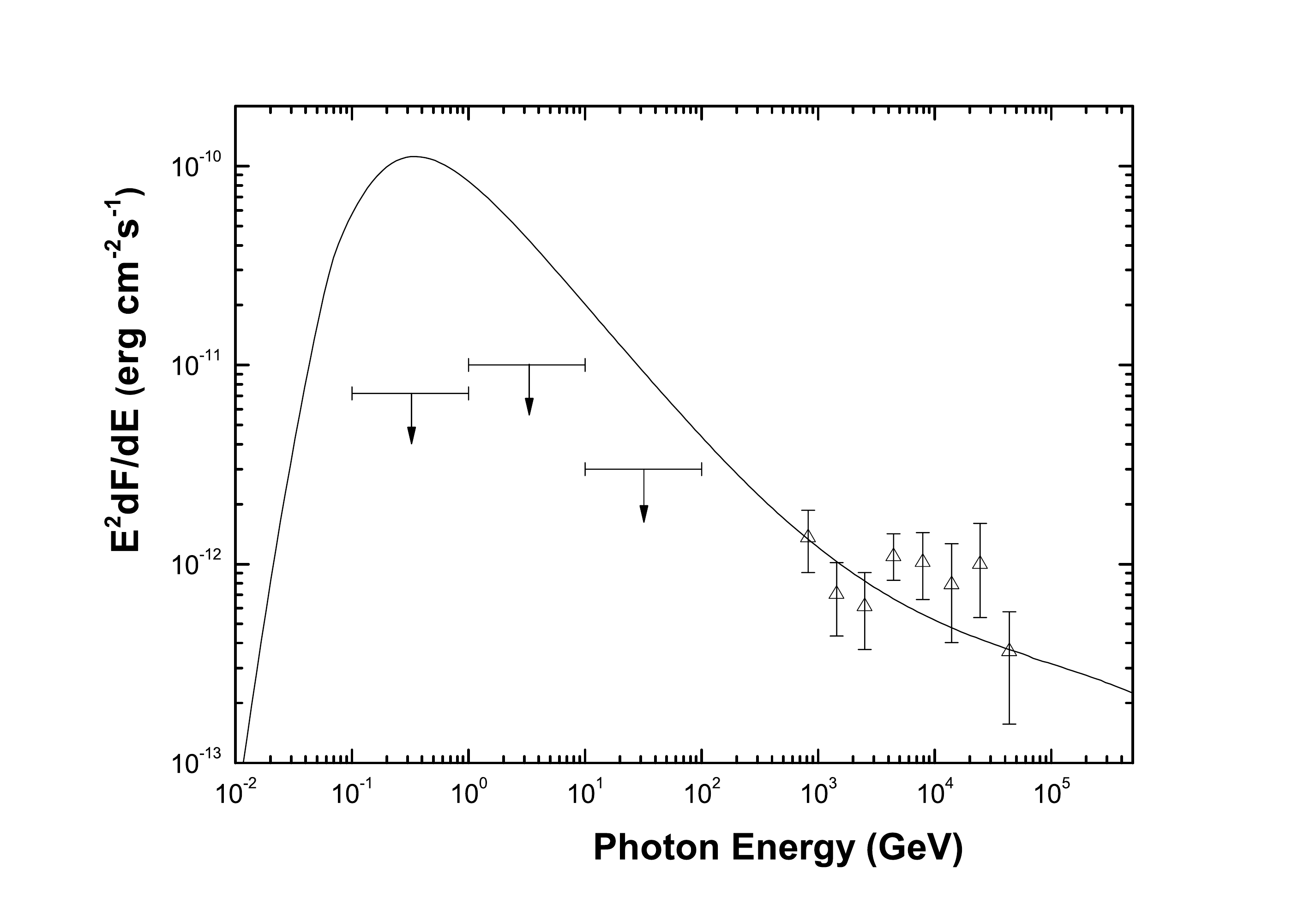}
 \caption{ Examples of the gamma-ray spectrum at the position of HESS J1858+020 resulting from the non point-like
 models of Table 3 but with an energy-independent diffusion.
From left to right, the two first panels show 
the cases with $\chi =0.001$ and 0.01. The third panel shows the case for with parameters $R$=30 pc, $p=2.17$, $
\delta=0$, $M_{cl}=2 \times 10^7$ M$_\odot$, and $\chi=0.01$.}
   \label{22}
\end{figure*}

{ We checked whether the case of an energy-independent diffusion, i.e.
$\delta=0$, can provide a good fit. To do that we have used 
the same 
parameters (except for $\delta$) as those shown in Table 3 to explore their matching to the multi-wavelength data.
As an example, results for $\chi=0.001$ and 0.01 are shown in Figure \ref{22}. Additionally, 
we have also set other parameters free (apart from fixing $\delta=0$)  to fit the H.E.S.S. data and the \fermi-LAT upper-limits: For instance, 
in order to fit the H.E.S.S. data with gamma-ray index $\sim 2.2$, the
proton spectrum index $p$ should also be 2.2 if $\delta=0$.
The right panel of Figure \ref{22}  shows the case for
$R$=30 pc, $p=2.17$, $
\delta=0$, $M_{cl}=2 \times 10^7$ M$_\odot$, and $\chi=0.01$, which matches 
the H.E.S.S. data, at the price of unavoidably violating
the \fermi-LAT upper-limits. The required MC mass would be much
larger than the one measured in the region, and the model is already ruled out by this.
A spectral break between GeV and TeV is needed to fit the spectral data, which
can be naturally produced by an energy-dependent diffusion model. }

\section{Discussion}

The GeV counterparts of some unidentified very high-energy sources
have been searched by \fermi-LAT and some of them show spectral breaks between the
GeV and the TeV band (e.g.\ HESS J1834-087/HESS J1804-216), which may imply
that the spectra cannot be treated as a single emission component
(Tam et al.\ 2010). These scenarios have been studied earlier by Funk et al. (2008), using EGRET data.
The very faint X-ray emission
of these unidentified
sources indicates that, unless evolved pulsar wind nebulae, these sources are likely not gamma-ray
emitters themselves, and 
cosmic-ray diffusion origins such as { the one} explored in this paper could be a possible alternative.

The cumulative diffusion model predicts concave gamma-ray spectra,
in which two peaks are present. The one at the lowest energies 
is attributed to the diffuse Galactic protons,
while the one at the highest energies, and putatively responsible for the
emission detected from HESS J1858+020 is attributed to the accelerated
protons that escaped from the nearby SNR G35.6-0.4.
{ It can be seen that the high-energy emission peak shifts to higher energies when the distance between the SNR and the cloud increases.}
Therefore, accelerator-illuminated gamma-ray
sources, which have the property of being TeV bright but GeV faint,
can be explained by diffusion effects. Such a
scenario was applied, for e.g., by Li \& Chen (2010) to explain the gamma-rays
of the four sources in the SNR W28 field. In W28, four gamma-ray sources
with various GeV and TeV brightnesses are accounted for by assuming different distances from the SNR center.
HESS J1858+020/SNR G35.6-0.4 might in principle be
another case of such generic phenomenology. 

In this paper we have studied HESS J1858+020 at GeV energies, 
while considering its possible association to the radio source, recently re-identified
as a SNR G35.6-0.4. The latter is found to be middle-aged ($\sim 30$ kyr) and has 
nearby MCs (Paron \& Giacani 2010). The \fermi-LAT First Source Catalog did not list any source coincident with the position of HESS J1858+020, but the closest \fermi-LAT source, 1FGL J1857.1+0212c, could have in principle been associated to it. A detailed analysis of 2 years of \fermi-LAT data { disfavors this association, and we have imposed upper limits} to the GeV emission from the region of interest.
Using both, a point-like and 3D models for the cosmic-ray injection and propagation out from the shell of the SNR, we consider whether the interaction between SNR G35.6-0.4 and the MCs nearby could give rise to the TeV emission of HESS J1858+020 without producing a GeV counterpart. 
We have found that although the phase space in principle allows for such a situation to appear, usual and/or observationally constrained values of the parameters (e.g., diffusion coefficients and cloud-SNR shell distance) would disfavor it.  Specifically, at the currently assumed distance of SNR G35.6-04 and the MC complex, the diffusion coefficient near this SNR should be greatly suppressed with respect to the average Galactic value, by more than one order of magnitude, in order for the gamma-ray phenomenology detected to be a viable outcome { of the three hadronic models considered here}. This may be  possible, and such cases have earlier been considered in the literature, but the size and density of this particular cloud may be too low  to generate such a slow diffusion timescale. 

\subsection*{Acknowledgments}

This work was supported by the  grants AYA2009-07391 and SGR2009-811, as
well as the Formosa Program TW2010005. AC is a
member of the Carrera del Investigador Cient\'ifico of CONICET, Argentina.
YC acknowledges support from the NSFC grant 10725312 and the 973
Program grant 2009CB824800. We thank 
 M. Lemoine-Goumard and S. Digel for comments. 
The \fermi-LAT Collaboration acknowledges generous ongoing support from a number of
agencies and institutes that have supported both the development and the operation of the LAT as
well as scientific data analysis. These include the National Aeronautics and Space Administration
and the Department of Energy in the United States, the Commissariat \`a l'Energie Atomique and
the Centre National de la Recherche Scientifique / Institut National de Physique Nucleaire et de
Physique des Particules in France, the Agenzia Spaziale Italiana and the Istituto Nazionale di Fisica
Nucleare in Italy, the Ministry of Education, Culture, Sports, Science and Technology (MEXT),
High Energy Accelerator Research Organization (KEK) and Japan Aerospace Exploration Agency
(JAXA) in Japan, and the K. A. Wallenberg Foundation, the Swedish Research Council and the
Swedish National Space Board in Sweden.
Additional support for science analysis during the operations phase is gratefully acknowledged
from the Istituto Nazionale di Astrofisica in Italy and the
Centre National d'\'Etudes Spatiales in France.

\label{lastpage}
\end{document}